\documentclass[conference]{IEEEtran}
\usepackage{cite}
\usepackage{hyperref}
\hypersetup{
    colorlinks=false,
}
\ifCLASSINFOpdf
   \usepackage[pdftex]{graphicx}
   \graphicspath{ {./images/} }
   \DeclareGraphicsExtensions{.pdf, .jpeg, .png, .jpg}
\else
\fi
\usepackage{pgfplots}
\usepackage{tikz}
\usepackage{adjustbox}
\usepackage{amsmath}
\usepackage{array}
\usepackage{url}
\begin{document}
\title{RouTEE: A Secure Payment Network Routing Hub using Trusted Execution Environments}

% use a multiple column layout for up to three different affiliations
\author{
    \IEEEauthorblockN{Junmo Lee}
    \IEEEauthorblockA{Seoul National University\\
    junmo.lee@snu.ac.kr}
\and
    \IEEEauthorblockN{Seongjun Kim}
    \IEEEauthorblockA{Seoul National University\\
    sms7002@snu.ac.kr}
\and
    \IEEEauthorblockN{Sanghyeon Park}
    \IEEEauthorblockA{Seoul National University\\
    lukepark@snu.ac.kr}
\and
    \IEEEauthorblockN{Soo-Mook Moon}
    \IEEEauthorblockA{Seoul National University\\
    smoon@snu.ac.kr}
}

% to anonymize author information
% \author{\IEEEauthorblockN{ \\}\IEEEauthorblockA{ \\ \\}\and}


%Telephone: (800) 555--1212, Fax: (888) 555--1212}


\maketitle

% \begin{abstract}
% The abstract goes here \cite{das2019fastkitten}, \cite{lind2019teechain}.
% \end{abstract}


% \fi
\IEEEpeerreviewmaketitle

% Page Limit and Formatting (from https://www.ieee-security.org/TC/SP2021/cfpapers.html)
% Submitted papers may include up to 13 pages of text and up to 5 pages for references and appendices, totaling no more than 18 pages. The same applies to camera-ready papers, although, at the PC chairs’ discretion, additional pages may be allowed for references and appendices. Reviewers are not required to read appendices.

% Papers must be formatted for US letter (not A4) size paper. The text must be formatted in a two-column layout, with columns no more than 9.5 in. tall and 3.5 in. wide. The text must be in Times font, 10-point or larger, with 11-point or larger line spacing. Authors are encouraged to use the IEEE conference proceedings templates. LaTeX submissions should use IEEEtran.cls version 1.8b. All submissions will be automatically checked for conformance to these requirements. Failure to adhere to the page limit and formatting requirements are grounds for rejection without review.

% text blocks
% As a general rule, do not put math, special symbols or citations in the abstract

\begin{abstract}

Cryptocurrencies such as Bitcoin and Ethereum have made payment transactions possible without a trusted third party, but they have a scalability issue due to their consensus mechanisms. 
Payment networks have emerged to overcome this limitation by executing transactions outside of the blockchain, which is why these are referred to as off-chain transactions. 
In order to establish a payment channel between two users, the users lock their deposits in the blockchain, and then they can pay each other through the channel. 
Furthermore, payment networks support multi-hop payments that allow users to transfer their balances to other users who are connected to them via multiple channels. 
However, multi-hop payments are hard to be accomplished, as they are heavily dependent on routing users on a payment path from a sender to a receiver. 
Although routing hubs can make multi-hop payments more practical and efficient, they need a lot of collateral locked for a long period and have privacy issues in terms of payment history.

We propose \textit{RouTEE}, a secure payment routing hub that is fully feasible without the hub's deposit. 
Unlike existing payment networks, RouTEE provides high balance liquidity, and details about payments are concealed from hosts by leveraging trusted execution environments (TEEs). 
RouTEE is designed to make rational hosts behave honestly, by introducing a new routing fee scheme and a secure settlement method. 
Moreover, users do not need to monitor the blockchain in real-time or run full nodes. 
They can participate in RouTEE by simply verifying block headers through light clients; furthermore, having only one channel with RouTEE is sufficient to interact with other users. 
Our implementation demonstrates that RouTEE is highly efficient and outperforms Lightning Network that is the state-of-the-art payment network. 

\end{abstract}

\section{Introduction}

Decentralized blockchains such as Bitcoin \cite{nakamoto2008bitcoin} and Ethereum \cite{wood2014ethereum} have been playing an important role by providing a payment infrastructure for mutually distrusting parties. 
But their lack of scalability is one of the biggest limitations to overcome.
Bitcoin, for instance, can execute a maximum of only 7 transactions per second, which is absolutely insufficient to replace existing payment systems. 
However, previous work \cite{gervais2016security} proved that it is not that effective to solve this inherent problem by simply changing parameters of Bitcoin (e.g., block size).

Payment networks such as Lightning Network \cite{poon2016bitcoin} and Raiden \cite{RaidenNetwork} have been proposed to process payments with off-chain transactions, achieving higher transaction throughput.
These transactions are not written in blockchains, so their performance is not limited by blockchain protocols. 
To execute \textit{off-chain} payments between two users, they open a payment channel by broadcasting a transaction (i.e., an \textit{on-chain} transaction) to the blockchain network and locking their deposits. 
If users are connected within the payment channel network, routing users between a sender and a receiver can convey the sender's balance to the receiver through their channels, and collect routing fees from the sender, which is called a \textit{multi-hop payment}.

However, payment networks suffer from various shortcomings. 
Once a payment channel is created, there is no way to withdraw some of the balance or to add additional deposits: users have to close the channel and open another one. 
Furthermore, payment channels should be settled at the latest balance state, but malicious users could try to settle channels at a past state. 
Constant surveillance is thus required for the underlying blockchain to prevent such attacks. 
In addition, various conditions must be satisfied to complete the multi-hop payment. 
Senders should keep tracking how the channel network topology changes to find proper payment paths to receivers. 
Intermediary routing nodes must be on-line, of course, and all channels on the payment path must possess a greater balance than the sender's payment amount.

A star graph with one central hub node is one of the most efficient topologies for payment networks \cite{avarikioti2020ride}. 
Users need only one channel with the hub and do not have to struggle to find payment paths. 
In addition, hubs would reduce the length of the multi-hop payment path and thus the total routing fees that a sender must pay. 
Despite this efficiency, these central nodes are impractical to employ. 
There is a privacy issue in that they can obtain most of the private information about payments that occurred in the network. 
Moreover, to deal with numerous transactions, hub nodes should lock a tremendous amount of deposits in their numerous channels.

A key idea in this work is to overcome these flaws by utilizing trusted execution environments (TEEs). 
An example of a TEE is Intel Software Guard Extensions (SGX) \cite{mckeen2013innovative, costan2016intel}, a prominent TEE product operated by extended instruction set architecture in recent Intel CPUs. 
SGX supports hardware-based memory encryption for application codes and data in memory in order to securely isolate them from adversaries. 
Sensitive data is stored in independent regions of memory, called \textit{enclaves}, which are inaccessible to other processes even with higher privilege levels.

We describe \textit{RouTEE}, a secure payment routing hub, providing a new payment network system. 
It makes multi-hop payments efficient and confidential, and does not require hosts to stake their assets. 
Users who have the channel with RouTEE can easily execute multi-hop payments via RouTEE, with less cost for verification. 
It is very cost-efficient because senders pay a routing fee only to RouTEE, and users do not have to open multiple channels, unlike existing payment networks. 
Due to SGX, the RouTEE protocol becomes concise and prevents rational hosts from misbehaving.

This work's main contributions are as follows: 

\begin{itemize}
    
    \item \textbf{Secure Payment Hub.} 
    Only one channel is enough to fully utilize RouTEE's features and payment details are securely protected by a TEE. 
    Basically, a user can open a channel with RouTEE through an on-chain transaction, similar to other payment networks. 
    But in RouTEE, there is a way to create the channel and get balance by executing off-chain transactions. 
    In addition, users can put additional deposits into their channels and withdraw part of their balance, which means that RouTEE provides high balance liquidity. 
    It's noteworthy that these channels do not contain any assets of RouTEE's host (i.e., only users' collateral), making it easy for multi-hop payments to succeed with fewer routing fees.

    \item \textbf{Less Burden on Users.} 
    RouTEE only requires users to run light clients to verify block headers, and this verification does not need to be processed in real-time. 
    Precisely speaking, users might need to check whether a specific block is included in the blockchain when other users want to pay them. 
    Furthermore, there is no additional data (such as penalty transactions) that users must store in other payment networks, and users execute multi-hop payments easily without knowing the network state (i.e., the network topology, the channel's balance state).

    \item \textbf{New Routing Fee Scheme.} 
    It is important not to let the host unexpectedly abort operating RouTEE as RouTEE manages a lot of deposits and should store them safely. 
    So we have introduced \textit{pending routing fees}, which motivate incentive-driven hosts to keep running RouTEE and settle users' balances properly. 
    In other words, hosts gain their entire pending routing fees only after every user has settled their whole balances.

    \item \textbf{Secure Settle Protocol.} 
    RouTEE collects all of its deposits together to make secure on-chain settlement transactions, which we call \textit{spend-all-settlement}. 
    In this way, RouTEE can prevent feeding fake chain attacks, where adversaries insert fake blocks not included in the main chain to deceive RouTEE as if they actually sent their assets. 
    \textit{spend-all-settlement} also improves anonymity since it can be regarded as a mixing service. 
    
\end{itemize}

We implement the RouTEE prototype using SGX for Bitcoin. 
We note that only one TEE is required to establish the RouTEE network. 
With only one RouTEE hub node, RouTEE can deal with more than 18,000 payments per second. 
Its throughput can be highly improved by batching payments.

\section{Background}

\subsection{Blockchains}

Since Bitcoin \cite{nakamoto2008bitcoin} appeared in 2008, there have been many other cryptocurrencies. 
In Bitcoin, which is the most famous example of \textit{proof of work} (PoW) \cite{dwork1992pricing} based blockchains, nodes are connected over peer-to-peer networks. 
Users make transactions when they want to pay other users and broadcast them with their cryptographic signatures to verify whether the owner of this asset has made this transaction or not. 
Then miners collect these transactions and try to solve a computationally expensive puzzle to make a valid block. 
A valid block means that it has a lower hash value than a target hash value. 
The target value is adjusted every 2,016 blocks (i.e., about two weeks) so it takes about 10 minutes on average to solve the puzzle. 
A block with a lower hash value than other blocks is considered to have a higher block difficulty. 
When a miner finds a valid block, it is broadcast to the blockchain network and appended to the chain of blocks that is append-only. 
Due to this hard work, blockchain data is difficult to revert.

When blocks with the same block number are made around the same time, it is described as a \textit{fork}. 
In Bitcoin, the \textit{longest chain rule} determines which block is valid. 
Each miner chooses a chain to continue mining, then one chain will become the longest chain and have the largest total difficulty, due to the difference in mining power for each chain. 
At that time, miners who were mining for other shorter chains move to the longest chain, and this chain becomes a \textit{main chain}.

Many cryptocurrencies, including Bitcoin, use an \textit{unspent transaction output} (UTXO) model. 
Transactions consist of inputs and outputs. 
Outputs represent an asset that can be spent freely by the owner. 
To make a transaction, senders select some of their unspent outputs as inputs of the transaction. 
Then they create new outputs for receivers, which means that the asset ownership has been moved to them. 
An output is locked with a lock called \textit{ScriptPubKey}, which can be unlocked with a key called \textit{ScriptSig} that only an owner of the output can create using its private key. 
These locks and keys are implemented in Bitcoin's Script language. 
Furthermore, all unspent outputs cannot be used more than once to prevent a double-spending attack (i.e., sending the same asset to several different receivers).

If a user wants to verify transactions, it has to run a full node to download all blockchain data including block headers and transactions, consuming plenty of network and storage resources. 
Then a user needs to verify every block header and transaction, which costs a lot of time and computation. 
A light client is suggested to deal with this problem. 
It only saves block headers, which require much less storage, and checks that they have actually solved PoW puzzles in a very short amount of time. 
Now users are able to verify that a certain transaction is included in a certain block with the block's header and the transaction's Merkle proof, which is called \textit{simplified payment verification} (SPV) \cite{nakamoto2008bitcoin}.

\subsection{Payment Networks}

Payment networks such as Lightning Network\cite{poon2016bitcoin} and Raiden \cite{RaidenNetwork} have emerged to solve the low scalability problem of blockchains. 
As on-chain payment transactions are very slow and expensive to execute, payment networks process payments privately with off-chain transactions through payment channels. 
This method massively improves their transaction throughput as their performance is not limited by the blockchain anymore. 
Payment networks also reduce the blockchain's storage because off-chain transactions are not written to blockchains. 
Only two transactions are recorded to open and close a payment channel (in normal situations where there is no dispute).

Users need to establish payment channels first to pay through payment networks. 
For instance, two users, Alice and Bob, make a funding transaction, which will lock their deposits, and broadcast it to peers in the blockchain network. 
Once they confirm that their funding transaction is included in the blockchain, they are considered to have opened a payment channel. 
Balances in the channel are determined by the amount of the deposit. 
Then, when they want to pay, they make a new off-chain payment transaction that changes their balance states and exchange their signatures for it.

If there are channels between Alice and Bob, and Bob and Charlie, then Alice cannot directly pay Charlie because there is no channel between them. 
To make such payments possible, payment networks offer multi-hop payments. 
A sender can execute multi-hop payments when a receiver is connected to the sender through multiple channels. 
For the above example, Alice first sends her balance to Bob with a routing fee for him. 
Next, Bob passes over the same amount of balance to Charlie through their channel. 
In this case, Bob is acting as a routing node for the multi-hop payment, and there could be more routing nodes. 
It should be noted that these channel state changes occur atomically, meaning that there is no circumstance where only some of the channels in the payment path are modified. 
This property is achieved by a Bitcoin's feature called \textit{hashed time-lock contracts} (HTLCs) in the Lightning Network.

When neither user (neither Alice nor Bob) wants to pay any longer, they can bilaterally settle their channel by broadcasting a settlement transaction that distributes their on-chain deposit to them immediately. 
Even if Alice is against closing the channel, Bob can unilaterally settle it by broadcasting any one of the off-chain payment transactions. 
In that case, Alice and Bob will be settled after a certain time limit if he broadcast the latest transaction. 
But if Alice detects Bob's malicious settlement attempts to broadcast a previous transaction rather than the latest one, she must broadcast the penalty transaction within the time limit in order to forfeit all channel deposits.

\subsection{Disadvantages of Payment Networks}

Even though payment networks can handle a lot more payments, other problems arise. 
Once a user's balance is exhausted, the user is no longer able to make payments through that channel. 
All the user can do is close the channel and create a new one, which incurs two expensive on-chain transaction fees. 
Likewise, withdrawing a fraction of the balance or transferring it to other channels is impossible, which implies that payment networks have low liquidity.

As mentioned above, any user can unilaterally close channels with the state at their own advantage. 
Therefore, users must keep downloading new blocks and watching all transactions to find malicious settlement transactions (i.e., running blockchain full node continuously). 
In other words, payment networks cost massive storage and network resources to achieve high scalability.

Although multi-hop payments seem convenient as they allow paying users without a direct payment channel, there are many restrictions. 
Users should broadcast and collect information about opening and closing channels to find out how payment networks are connected. 
Then, they need to search for proper payment paths in which all channels have a balance greater than the payment amounts. 
Unfortunately, to protect privacy, channel balance information is not provided to other users, so it is hard to know which path could complete the multi-hop payment. 
There is no choice but to try the multi-hop payment through various paths until it succeeds. 
Users may have to split the payment amount within several multi-hop payments in order to complete the payment. 
Moreover, considering that the mean shortest payment path length is about three in the Lightning Network \cite{seres2020topological}, and that a multi-hop payment through the shortest path could fail, senders might end up paying for two routing nodes on average, or even far more. 
We note that users open seven channels on average in the Lightning Network \cite{seres2020topological}, which means that users pay a lot of channel creation costs and that they need to split their assets.

Moreover, there are privacy and security issues. 
Recent research \cite{herrera2019difficulty, kappos2020empirical} shows that there is a way to figure out how much balance each user in the channel has, which can lead to payment information leakage (e.g., who the sender or receiver was, which routing nodes were used, how much amount to pay). 
(e.g., who was the sender, the receiver, or routing nodes, how much amount to pay). 
A \textit{wormhole attack} \cite{malavolta2019anonymous} allows two attackers on a payment path to intercept routing fees for other routing nodes between them. 
In addition, it has been proven that adversaries are able to interrupt other routing nodes to monopolize routing requests, by holding the balances of other nodes for a long period \cite{perez2020lockdown} and exhausting their channels deliberately \cite{rohrer2019discharged}.

\section{Design}

\subsection{Design Overview}

RouTEE performs as a payment routing hub which makes multi-hop payments more practical, keeping private data confidential within SGX. 
RouTEE can support any UTXO-based blockchain such as Bitcoin or Litecoin as it only utilizes UTXO's features and simple transactions that just transfer coins.

There is a set of users who participate in RouTEE, and a host who runs the RouTEE platform using an SGX-enabled machine and feeds blockchain data into SGX. 
We note that users are able to verify that the RouTEE program is actually executed in SGX and build secure communication sessions through the \textit{remote attestation} properties of SGX \cite{costan2016intel, anati2013innovative}. 
First, SGX makes a \textit{measurement} which consists of program codes, the enclave state, and additional data (e.g., a public key to establish secure sessions). 
Then it signs the measurement with its private key concealed within it. 
After that, users can verify this signature through Intel's online attestation service, having confidence that the signature is truly created by SGX and that it is executing the correct RouTEE program.

It is important to separate channel creation from deposit locking in RouTEE, meaning that users can build channels even with off-chain transactions that do not require expensive on-chain transaction fees for blockchain miners. 
To be more specific, like existing payment networks, users can open channels and add balance at the same time via on-chain transactions, or they can simply open channels that contain no balance through off-chain transactions. 
This separation also allows users to add more deposits to their channels or withdraw their balance partially without the channel closing.

We also note that RouTEE makes multi-hop payments way more achievable. 
For users, tracking the network topology is no longer necessary to find proper payment paths, since every user is connected via RouTEE. 
This implies that RouTEE is a unique routing node and that senders have fewer routing fees to pay for their multi-hop payments. 
Moreover, users can determine whether their multi-hop payments are valid or not with a simple block header verification. 
For RouTEE, it just delivers the senders' assets to receivers, meaning that RouTEE's asset is not required. 
Accordingly, there is no case where multi-hop payments fail as RouTEE has not enough balance, and the host can operate RouTEE without any collateral.

% insert figure
\begin{figure}[t]
    \centering
    \includegraphics[width=\linewidth]{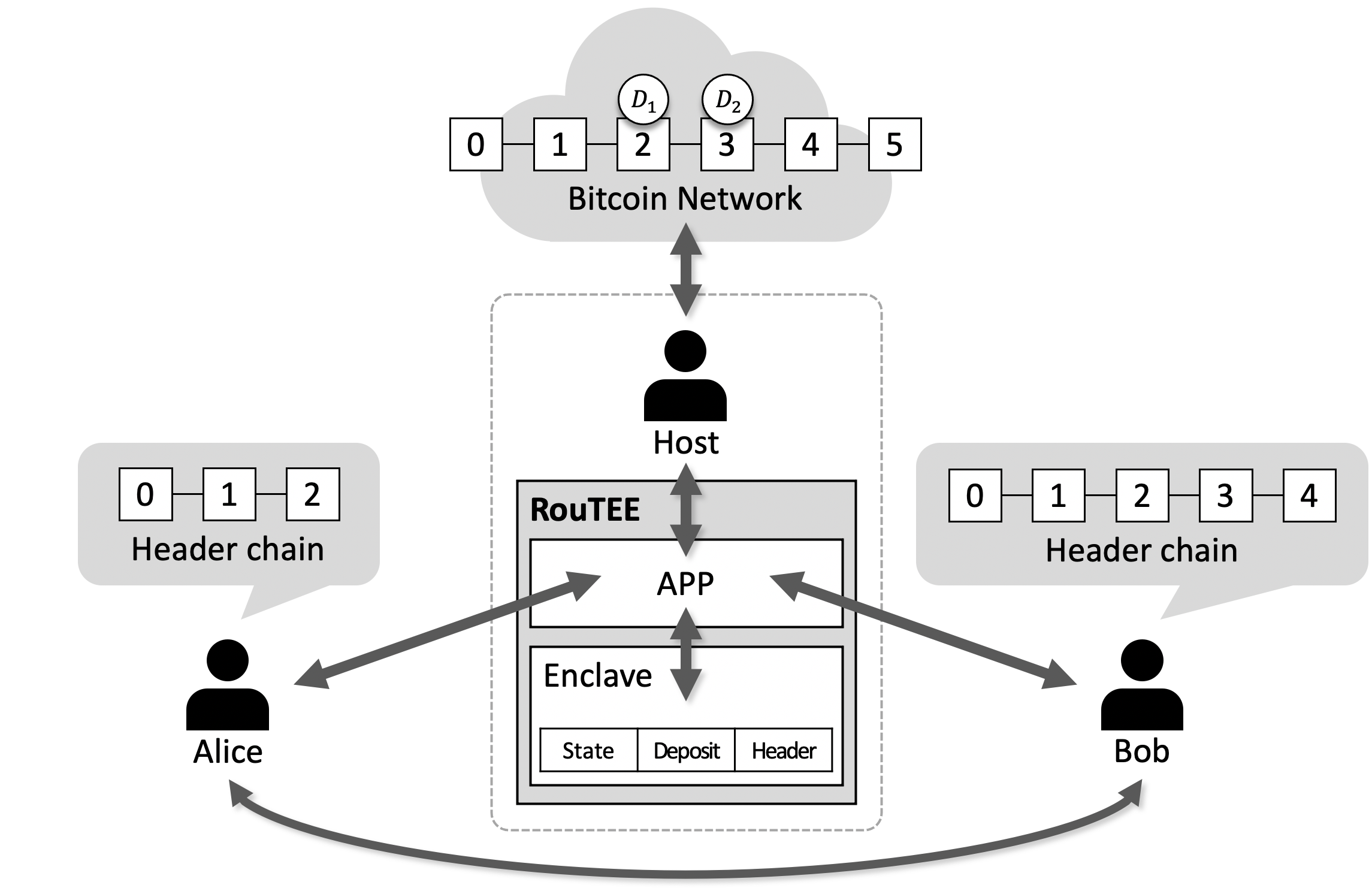}
    \caption{
        The RouTEE system overview. 
    }
    \label{fig:overview}
\end{figure}

Fig.~\ref{fig:overview} describes RouTEE's overall system. 
The host first initializes RouTEE by feeding blockchain data until it reaches the latest block. 
Then users must verify that RouTEE is initialized successfully within SGX and establish secure sessions through remote attestation. 
All important data is kept safe inside SGX's enclave. 
The host and users interact with SGX through an interface process that is not protected by SGX. 
This means that the host has full controls of the process and can manipulate it. 
Users should maintain their own header chains, but their perspectives on the blockchain (e.g., what is the latest block number, which block is correct) might be different.

\begin{figure}[t]
    \centering
    \includegraphics[width=0.9\linewidth]{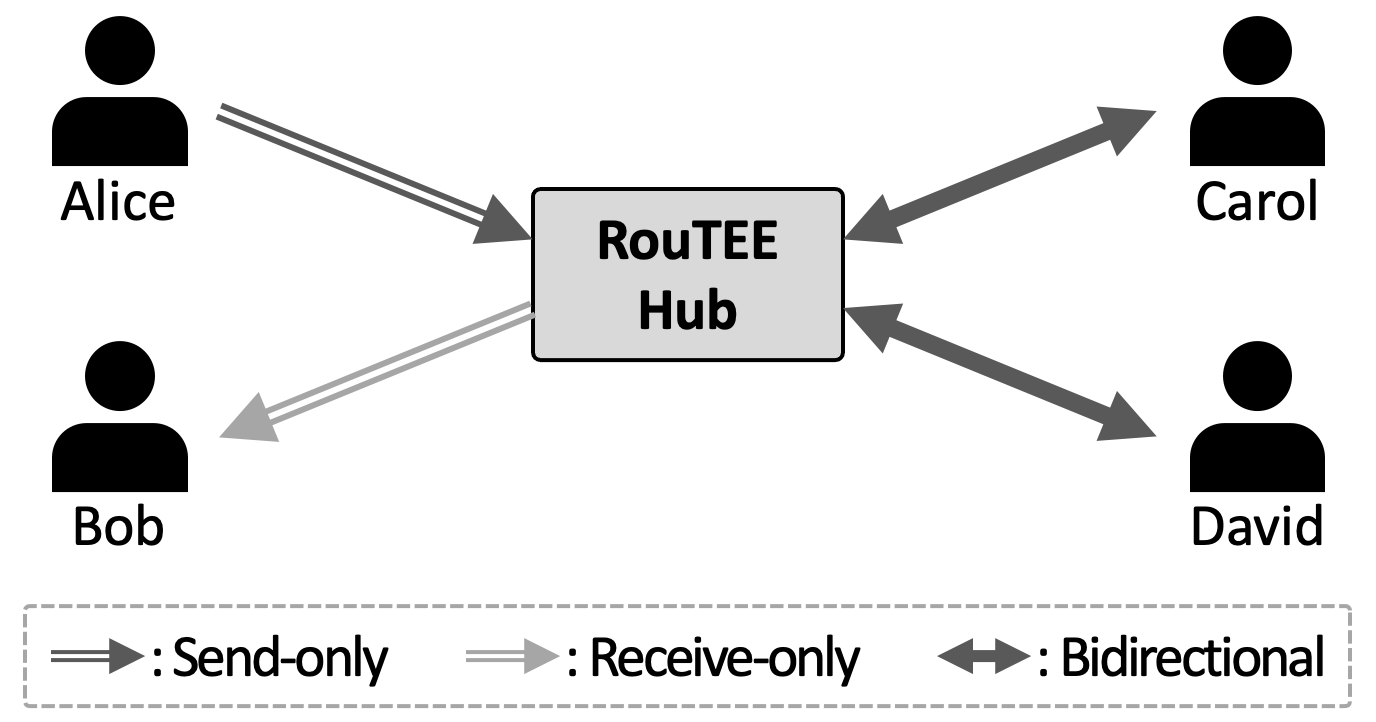}
    \caption{
        An example of the RouTEE network. 
        It forms a star graph which has RouTEE as a unique intermediary node. 
        Each user has one of three kinds of channels with RouTEE, but channels between users are not necessary. 
    }
    \label{fig:topology}
\end{figure}

In order to interact with RouTEE, users need one of the three types of channels: \textit{send-only channels}, \textit{receive-only channels}, and \textit{bidirectional channels}. 
There being no balance in RouTEE at first, users who want to be multi-hop payment payers broadcast on-chain \textit{deposit transactions}, opening send-only channels and gaining balance in their channels. 
Deposit transactions will be included in some blocks, which we call \textit{source blocks}. 
These transactions are automatically applied to user states inside SGX, as the host should insert every newly created block and transaction. 
Users who are going to be payees set their \textit{boundary blocks} by simple off-chain transactions, building receive-only channels. 
A boundary block is the latest block among blocks that a user believes valid. 
That is, users cannot receive from other users who have balance derived from newer source blocks than their boundary blocks. 
To sum up, users with more than a 0 balance and users who set their boundary blocks can be considered as having send-only channels and receive-only channels, respectively. 
Users who meet both conditions have bidirectional channels. 
Fig.~\ref{fig:topology} shows RouTEE's network example with various types of channels.

When both the payer and the payee are ready, multi-hop payments can be executed. 
Senders pay RouTEE \textit{pending routing fees} for their multi-hop payments. 
Pending means that these routing fees are not yet in the host's possession. 
To prevent the host from abnormally quitting RouTEE operations, pending routing fees are confirmed only after users settle their balance successfully.

Users can receive on-chain assets (e.g., BTC in Bitcoin) by settling their balances in RouTEE. 
RouTEE makes \textit{settlement transactions} that use all deposits it owns inside SGX to securely settle a batch of users who requested a settlement. 
This is called \textit{spend-all-settlement}, which effectively protects on-chain assets from misbehaving hosts who try to steal them by exploiting their fake deposits. 
Then rational hosts will honestly broadcast settlement transactions to the blockchain network to confirm their pending routing fees.

% insert figure

\subsection{Adversary model and assumptions}

Using TEEs is essential to building the RouTEE system. 
SGX is one of the well-known products of a TEE, but it is not perfect in terms of security. 
Recent works revealed several vulnerabilities of SGX \cite{brasser2017software, lee2017hacking, van2018foreshadow, biondo2018guard, chen2019sgxpectre}. 
This means that, in practice, some of SGX's properties such as integrity and confidentiality might not be guaranteed. 
However, a great deal of research has been conducted to surmount these limitations \cite{seo2017sgx, shih2017t, chen2017detecting, oleksenko2018varys, ahmad2019obfuscuro}, and any other TEE (e.g., ARM TrustZone \cite{alves2004trustzone}) can also be employed to implement RouTEE. 
So in this paper, we assume that there is no security issue with SGX itself.

Since RouTEE is operated atop of the blockchain, some basic properties of the blockchain should be satisfied. 
First, any valid transactions with a proper amount of transaction fee are included within the reasonable upper-bounded time period. 
Though adversaries could bribe miners not to contain certain transactions in their blocks, it must cost a lot to keep miners corrupt for a long period. 
Second, blocks with more than \textit{k-confirmations} are not removed from the blockchain. 
In other words, a block followed by more than \textit{k}-1 blocks can be considered immutable. 
\textit{k} can be any integer bigger than 0, but commonly \textit{k} is 6 in Bitcoin. 
This assumption also indicates that there is no attacker who can change the main chain of the blockchain with tremendous hash power.

We assume all participants are rational and incentive-driven. 
Their goal is to maximize their own interests. 
This assumption is also applied to adversaries (i.e.,\textit{ rational adversaries} \cite{dong2017betrayal}), so they never choose strategies that result in financial losses for themselves, even if the strategy brings greater losses to victims. 
There might be \textit{byzantine adversaries} \cite{shostak1982byzantine} who do not care about their payoff and act irrationally (e.g., the host could turn off the RouTEE's power, abandoning all its accumulated pending routing fees). 
However, they are quite unrealistic so we are not concerned about them. 
If we attempt to mitigate this kind of vulnerability of a hub struct, we could separate the hub into several nodes, backup every payment and state, and introduce more complicated protocols. 
Still, this dilutes the benefits of the hub, which implies that there is a trade-off between security and performance. 
So we concentrate on inducing rational adversaries to behave honestly.

\subsection{Design Challenges}

\textbf{Misbehaving Host.} 
There are two types of participant in RouTEE: a user and a host. 
The host has more authority and attack strategies than users, considering that it can directly access RouTEE and be the user. 
For example, the host is actually located between RouTEE and the users, passing messages between them. 
Although the host cannot find out message contents as messages are encrypted, the host can deliberately delay or drop them. 
Furthermore, blockchain data is important for operating RouTEE correctly, and the host plays the role of data feeding by running a blockchain full node client. 
Though basic block verification is performed inside RouTEE's SGX, the host with moderate mining power and sufficient time can make and insert a fake blockchain because there is no other chain in SGX to compare for determining which chain is valid based on the longest chain rule. 
That is, the host can deceive RouTEE as if he had staked the deposit and can get invalid balance inside RouTEE. 
Then by paying or settling the balance, the host will gain illegal profits. 
Executing all full node codes inside the SGX could be the solution, but it increases the size of the \textit{trusted computing base} (TCB) too much, making the solution impractical as it would provide various attack vectors to adversaries. 
Besides, if aborting RouTEE costs nothing (i.e., there is no asset of the host inside RouTEE), the rational host could quit running RouTEE, and unsettled balances inside RouTEE would totally disappear. 
To protect RouTEE from irresponsible hosts, we have introduced an appropriate incentive model to make honesty the best strategy along with secure protocols to prevent illegal payments and settlements.

\textbf{Verifying Blockchain Data.} 
In existing payment networks, all users must observe every transaction to prevent invalid settlements due to their mutual distrust. 
However, it is a huge waste running full node clients to detect illegal attempts, because blockchain read/write operations are expensive and most transactions are irrelevant to users. 
To resolve this inconvenience, we leverage SGX and light clients. 
Since payments and settlements are performed securely inside SGX, no one can manipulate user states or produce invalid settlement transactions. 
Thus, users do not have to keep watching the blockchain in real-time. 
Instead, light-weight verification with block header data from light clients can efficiently preclude malicious payments. 
All users have to do is simply verify block headers and update their boundary blocks, before they become multi-hop payment payees. 
We emphasize that this process is not required for every payment.

% \clearpage

\section{Protocol} 
This section describes RouTEE protocol details, including how hosts initialize RouTEE and feed blockchain data, along with how users create channels, add deposits, execute multi-hop payments, and settle their balances.

\subsection{Initialization}

There are a few things that the host has to do to start RouTEE.
The host must register the host's public key to prevent users from executing host-only operations such as inserting Bitcoin blocks. 
Then the host sets a minimum amount of routing fee per multi-hop payment. 
Other users cannot modify the amount since that operation needs the host's cryptographic signature.

Next, block headers need to be inserted inside SGX until it reaches the latest block. 
Although the header chain verification is a much lighter process than the full blockchain verification, it takes a lot of time to check every block from the genesis block (e.g., In Bitcoin, there are about 655,000 blocks as of November 2020). 
To shorten this procedure, it can start from one of the \textit{checkpoint blocks}, which are hard-coded into standard blockchain clients and widely accepted as valid blocks. 
Likewise, the host can choose any one of the recent blocks as a start block and hard-code it into the RouTEE's source code. 
In this case, the verification process becomes incredibly simple. 
However, to filter out fake blocks with a too low difficulty, the start block should be more than 2,016 blocks far from the latest block in order to calculate Bitcoin's block difficulty inside the SGX.

% variables
\newcommand{\txAvgFee}{fee_{avg}}

Lastly, to measure the average on-chain transaction fee per byte (i.e., $\txAvgFee$), the host inserts on-chain transactions included in some recent blocks (e.g., for 2,016 blocks, the same as the block difficulty change period). 
Bitcoin's block headers contain the Merkle root of transactions, hence RouTEE can determine that these inserted transactions are actually involved in a certain block with its header. 
Based on this average transaction fee, RouTEE sets the amount to charge users for on-chain settlement transaction fees.

All block headers inserted since the initialization should be stored inside the SGX. 
However, we note that this block header chain occupies little storage because Bitcoin's block header size is just 80 bytes. 
Even if there were 1,000,000 blocks in Bitcoin, for example, it would not exceed 80 MB.

\subsection{Channel Creation}

After the initialization, users are able to obtain the result of the header chain verification (e.g., the initial block to start verifying and the latest block inside SGX) and build secure sessions with RouTEE by remote attestation. 
If they believe the initialization was successful, they will participate in RouTEE. 
Due to the session, the content of messages between the user and RouTEE is not revealed to others, especially the host.

% insert figure
\begin{figure}[t]
    \centering
    \includegraphics[width=\linewidth]{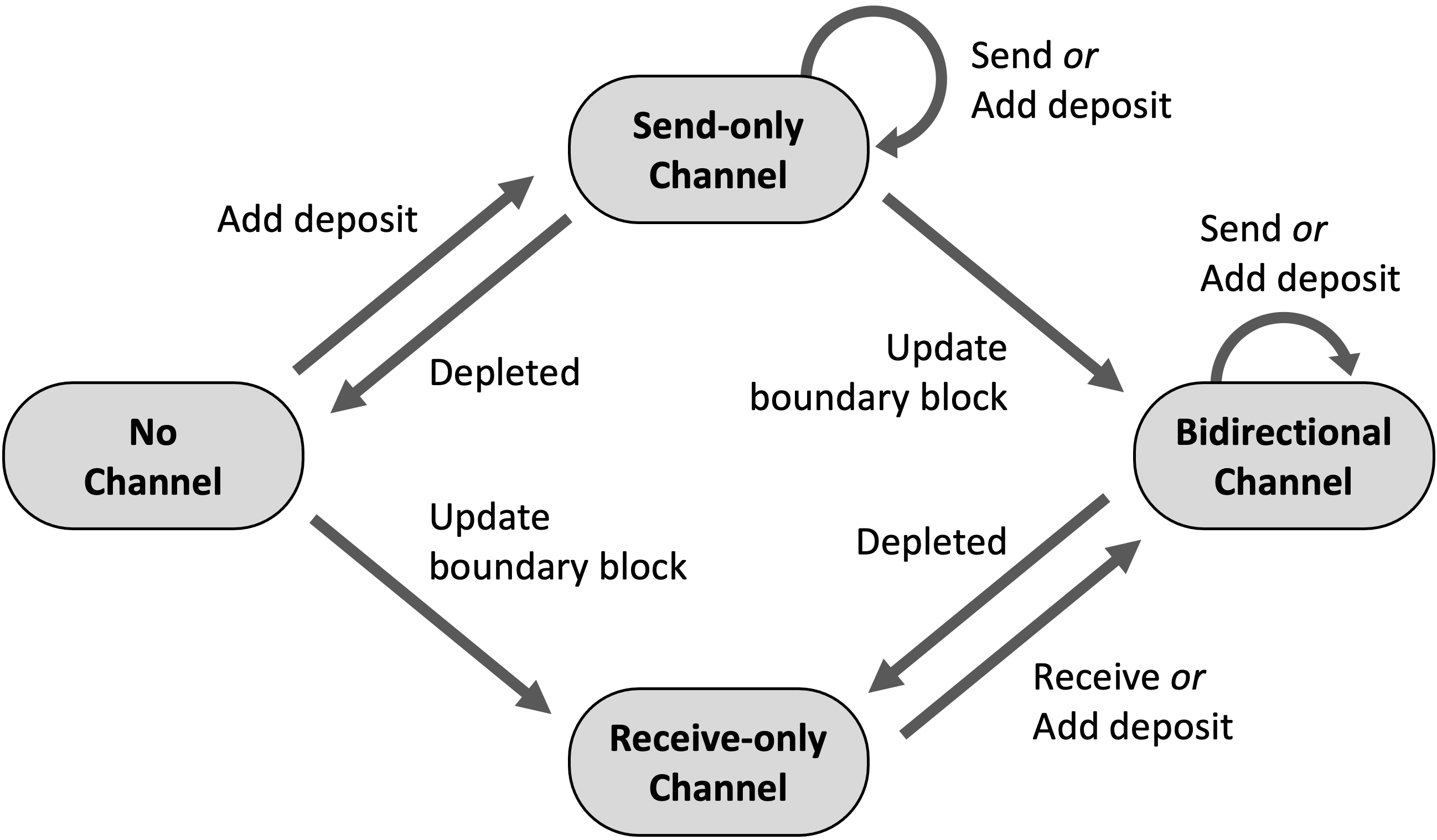}
    \caption{
        RouTEE channel type transition diagram.
        Channels could be dynamically opened, closed, and changed to other types of channels.
    }
    \label{fig:channel_transition}
\end{figure}

As we mentioned in Section III-A, there are three kinds of channels in RouTEE, and users need to create their channels first. 
In fact, channels in RouTEE are closer to the users' state than channels that exist explicitly in other payment networks. 
Thus, a channel's state changes dynamically as a user's state (i.e., balance, boundary block) changes. 
Fig.~\ref{fig:channel_transition} shows channel state transitions through several user operations. 
Users who have no channel at first can only execute \textit{add\_deposit} or \textit{update\_boundary\_block} operations to open initial unidirectional channels.

% insert figure
\begin{figure*}[t]
    \centering
    \includegraphics[width=\linewidth]{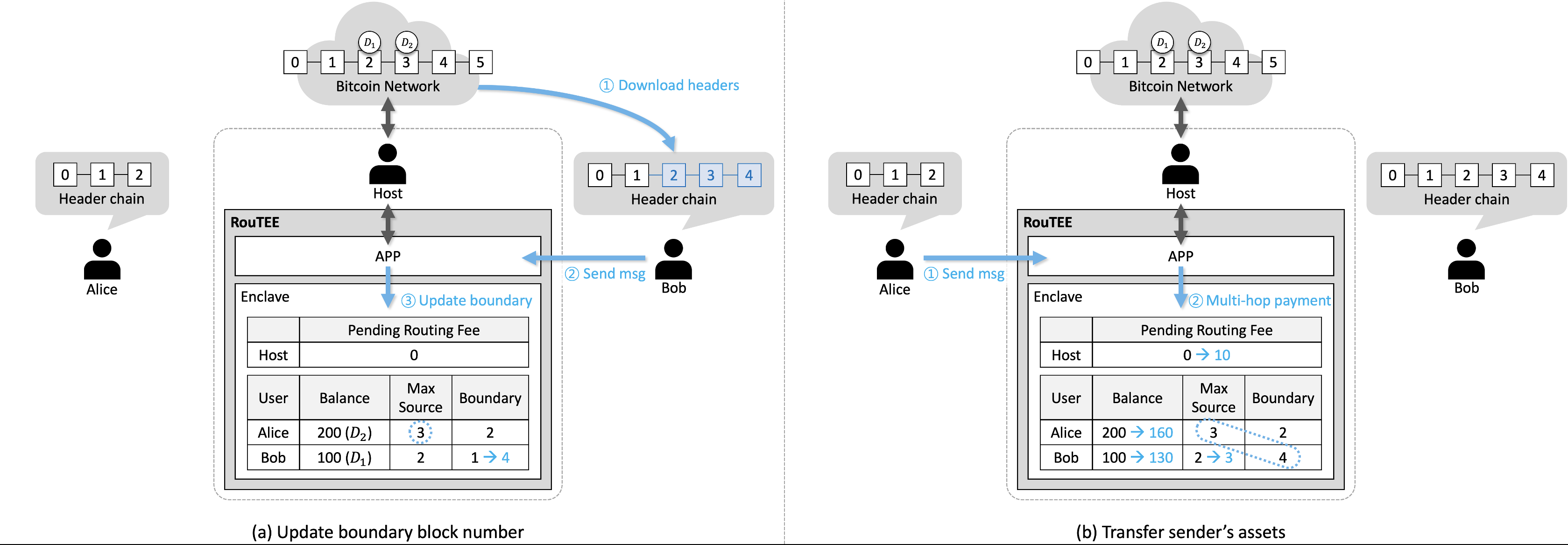}
    \caption{
        An example of a multi-hop payment in RouTEE. 
        Alice tries to send her balance derived from the deposit $D_2$ in block 3 to Bob, but his boundary block number is 1 which is less than 3. 
        This means that Bob does not trust Alice's balance yet. 
        (a) Alice informs Bob of her maximum source block number (in this case, 3). 
        Bob downloads block headers from full nodes to verify whether block 3 is valid. 
        He receives up to block 4 and sets his boundary block as it by \textit{update\_boundary\_block}, in order to prepare to receive from Alice. 
        (b) Alice executes a multi-hop payment to send 30 to Bob. 
        RouTEE checks that Bob believes Alice's balance is valid (i.e., Alice's maximum source block number is less than or equal to Bob's boundary block number. In this case, check $3 \leq 4$). 
        Then Alice's balance is moved to Bob, and the host gets a pending routing fee for this multi-hop payment from Alice. 
        Now Bob has balances from different source blocks, so Bob's maximum source block number should be updated to 3. 
    }
    \label{fig:multi-hop_payment_example}
\end{figure*}

\subsubsection{Add user}

Before users create channels, they need to enroll in RouTEE first via \textit{add\_user}. 
The user provides its public key, \textit{user address}, and \textit{settle address} to RouTEE. 
The user's public key is utilized to verify the user's signature when the user executes user operations such as multi-hop payments. 
A user address acts as the user's ID in RouTEE. 
When the user settles its balance later, on-chain assets are shifted to its settle address. 
The user should inform its settle address in advance since the user's channel might be settled automatically without the user's request (See section V-B, VII-C for more details). 
Then RouTEE stores this user information in a state, and this user is then ready to build unidirectional channels through \textit{add\_deposit} or \textit{update\_boundary\_block}. 
We note that each user has a \textit{nonce} field in a user state in RouTEE. 
User operations described below only accept messages with the value equal to the nonce, increasing the nonce by one and thus invalidating the messages.

\subsubsection{Add deposit}

\textit{add\_deposit} is a preregistration process to inform RouTEE of a user who is going to lock its deposit. 
The user sends a beneficiary address, namely its user address, to RouTEE. 
Then RouTEE generates a new random address, called a \textit{manager address}, inside SGX to receive the user's on-chain deposit and transmits a response message containing the manager address. 
RouTEE also saves these addresses to a \textit{pending deposit list} to detect a user's deposit transaction later. 
Lastly, the user sends on-chain coins via a \textit{pay-to-public-key-hash} (P2PKH) transaction, one of the standard transactions in Bitcoin, from any sender's addresses to the manager address.

% variables
\newcommand{\depositAmount}{D_{amount}}
\newcommand{\balanceIncreaseAmount}{B_{increase}}

The source block, which includes the deposit transaction, will be fed to RouTEE by \textit{insert\_block} operation, which deals with deposit transactions, updating users' maximum source block numbers and their balances. 
But the increased amount of the balance is less than the deposit amount. 
Since this on-chain deposit, namely a UTXO in Bitcoin, will be used to make a settlement transaction, RouTEE collects a fare from the user's balance in advance for the settlement transaction fee. 
Let $\depositAmount$ be the amount of this deposit transaction and $\balanceIncreaseAmount$ be the balance increase amount. 
Then $\balanceIncreaseAmount$ is determined as: 
$$ \balanceIncreaseAmount = \depositAmount - 148 \cdot \txAvgFee $$
This is because an on-chain transaction fee is proportional to the size of the transaction and a size of one input of P2PKH transaction is 148 bytes (See section IV-D for more details).

As described in Fig.~\ref{fig:channel_transition}, users who have no channel gain send-only channels, and users with receive-only channels obtain bidirectional channels by \textit{add\_deposit}. 
If a deposit transaction has not appeared in a certain block time period (e.g., for 100 blocks), RouTEE concludes that the user does not broadcast the deposit transaction, and deletes information about the deposit from the pending deposit list.

\subsubsection{Update boundary block}

\textit{update\_boundary\_block} increases a user's boundary block number and is performed as follows. 
In order to get valid block headers, users can download header chains from full node users by running light clients. 
If they trust the data source, it is sufficient to receive only one block and select it as a boundary block. 
Then the user sends its user address, signature, and boundary block's block number and hash value to RouTEE. 
The signature is required to authenticate the user. 
If the user's boundary block parameters match a header within RouTEE's header chain, and the new boundary block's number is greater than the user's previous one, RouTEE updates the user's boundary block number to it. 
As illustrated in Fig.~\ref{fig:channel_transition}, for users who have no channel with RouTEE, \textit{update\_boundary\_block} permits them to receive valid balances from other users, opening receive-only channels. 
This operation also opens bidirectional channels for users with send-only channels.

\subsection{Payment}

There are three agents who participate in a multi-hop payment: a sender, a receiver, and RouTEE as a unique intermediary. 
We note that, in RouTEE, there is no direct channel between two users and every payment is always a 2-hop payment, which is the shortest and most efficient multi-hop payment. 
As described in Fig.~\ref{fig:multi-hop_payment_example}, multi-hop payments proceed in two phases: a \textit{ready phase} and a \textit{transfer phase}. 
Suppose that the sender already executed \textit{add\_deposit} and has some balance to send. 
During the ready phase, the receiver should be ready to receive a balance from the sender. 
If the receiver's boundary block number is less than the sender's maximum source block number, the receiver cannot fully trust the sender's balance, because that means the sender might have invalid balances from source blocks that the receiver does not consider valid yet. 
In order to allow the sender to execute a multi-hop payment, the receiver should update its boundary block to the block that is newer than the sender's maximum source block by executing the \textit{update\_boundary\_block} operation. 
But if the receiver already has a proper boundary block number, this ready phase can be omitted.

After both users are ready, they proceed to the transfer phase. 
The sender transmits a message to RouTEE, including the sender's address, its signature, the receiver's address, a payment amount, and a routing fee amount. 
The sender's address and signature are used to authenticate the sender. 
Then inside SGX, RouTEE checks conditions: whether the sender has enough balance to afford this payment, and whether the sender's maximum source block number is equal to or greater than the receiver's boundary block number. 
Once the payment request is deemed to be acceptable, the receiver gets paid and RouTEE obtains a pending routing fee from the sender. 
However, RouTEE cannot settle this pending fee now. 
It will be confirmed after the settlement is successfully completed (See section IV-E for more details). 
Then the receiver's maximum source block number should be updated because the receiver may have the balance from a newer source block after this payment.

There are some notable observations in the payment process. 
A channel between users is not necessary, so they can save a lot of channel creation costs, which entail expensive on-chain transaction fees. 
Payments could succeed even with receivers being off-line, and only receivers may need to read block headers to update boundary blocks. 
Furthermore, when receivers should do so, they do not have to synchronize with the blockchain until it reaches the latest block. 
After updating boundary blocks, users have no need to care about events in the blockchain (i.e., not monitoring new transactions). 
Users can set their own \textit{k} value higher than RouTEE's for the \textit{k-confirmation} assumption, meaning that they can be as conservative as they want when judging a block's finality. 
Also, senders do not have to consider the network state (e.g., the topology, balances in channels, and payment paths).

\subsection{Settlement}

Users can request to settle balances in their channels for any amount. 
As a settlement requires an on-chain transaction, users need to pay settlement fees for making an on-chain settlement transaction. 
Settlement protocol details are as follows. 
A user sends the balance amount to settle, the settlement fee to be paid, its address and its signature, to RouTEE. 
To prevent free rides on the settlement, a minimum amount of settlement fee has been put in place. 
Users need to pay more than $34 \cdot \txAvgFee$, because the size of one output of a P2PKH transaction is 34 bytes. 
Then RouTEE verifies that the user has enough balance for this settlement request, and inserts it into the settlement request queue in which requests are sorted by the amount of the settlement fee.

% variables
\newcommand{\txInputNum}{Tx_{inputs}}
\newcommand{\txOutputNum}{Tx_{outputs}}
\newcommand{\txSize}{Tx_{size}}
\newcommand{\txFee}{Tx_{fee}}

Now RouTEE tries to make a settlement transaction, which settles the largest number of users with sufficient settlement fees, using a greedy method. 
Let $\txInputNum$ be the number of transaction inputs, and $\txOutputNum$ be the number of transaction outputs. 
Then, the size of a P2PKH transaction (i.e., $\txSize$) can be calculated as: 
$$ \txSize = 148 \cdot \txInputNum + 34 \cdot \txOutputNum + 10 $$
10 bytes is for other extra fields of a transaction, excepting inputs and outputs. 
Then an on-chain transaction fee (i.e., $\txFee$) can be determined as:
$$ \txFee = \txSize \cdot \txAvgFee $$
The important thing is that a settlement transaction uses all deposits (i.e., UTXOs) that RouTEE owns, which is a \textit{spend-all-settlement} method. 
Thus a settlement transaction's $\txInputNum$ is the number of deposits, and $\txOutputNum$ is the same as the number of users who requested a settlement plus one to make a \textit{leftover deposit} with leftover balances after the settlement. 
Users who participate in the settlement have to split the fee for this leftover deposit. 
As mentioned in section IV-B, deposit senders have already paid for this settlement. 
Briefly, if the amount of collected fees from users (i.e., from deposit senders and settle requesters) is greater than $\txFee$, RouTEE creates a settlement transaction and broadcasts it. 
Otherwise, it just waits for other settle requests with sufficient fees. 
However, if $\txAvgFee$ soars, settle request users should pay more fees because they need to pay for the settlement transaction's inputs also (as pre-collected fares become not enough).

% variables
\newcommand{\routingFeeToBeConfirmed}{RF_{confirmed}}
\newcommand{\pendingRoutingFee}{RF_{pending}}
\newcommand{\settleAmount}{S_{amount}}
\newcommand{\totalBalance}{B_{total}}

After making a secure settlement transaction, some portion of the pending routing fee will be confirmed for the host. 
Let $\pendingRoutingFee$ be the amount of total pending routing fees, $\settleAmount$ be the amount of balances to be settled by this settlement transaction, and $\totalBalance$ be the amount of total user balances in RouTEE. 
Then the amount of pending routing fees to be confirmed (i.e., $\routingFeeToBeConfirmed$) is determined as: 
$$ \routingFeeToBeConfirmed = \pendingRoutingFee \cdot \frac{\settleAmount}{\totalBalance} $$
This means that the host can gain the whole routing fees only when the host settles all users' balances completely, that being when $\settleAmount$ is equal to $\totalBalance$. 
The settlement transaction and $\routingFeeToBeConfirmed$ are stored and will be used in the \textit{insert\_block} operation to finally confirm routing fees and allow the host to withdraw that confirmed routing fee.

\subsection{Block Insertion}

A host needs to insert block headers and transactions into RouTEE to detect deposit transactions and settlement transactions. 
\textit{insert\_block} operates as follows. 
The host inserts a block header and its transactions into RouTEE. 
This operation can only be performed by the host as it requires the host's signature also. 
In order to verify block data, RouTEE verifies that the new block header satisfies the PoW conditions and that the Merkle root can be reproduced with these transactions. 
It then adds this header to the header chain, looks at transactions to update $\txAvgFee$, and searches for deposit transactions and settlement transactions. 
If a deposit transaction is found, it gets the corresponding user address with this transaction from the pending deposit list, increases the user's balance by $\balanceIncreaseAmount$, and updates the user's maximum source block number to this block number. 
If there is a settlement transaction, it similarly brings the matching $RF_{confirmed}$ value and increases the host's balance (i.e., confirmed routing fees) by that value.

% \clearpage

\section{Security Analysis} 

In this section, we show various kinds of attack strategies that hosts could choose and explain how the RouTEE protocol protects user balances from rational hosts. 
We emphasize that there is no case in which hosts gain unfair benefits while users lose their balances.

\begin{figure*}[t]
    \centering
    \includegraphics[width=0.85\linewidth]{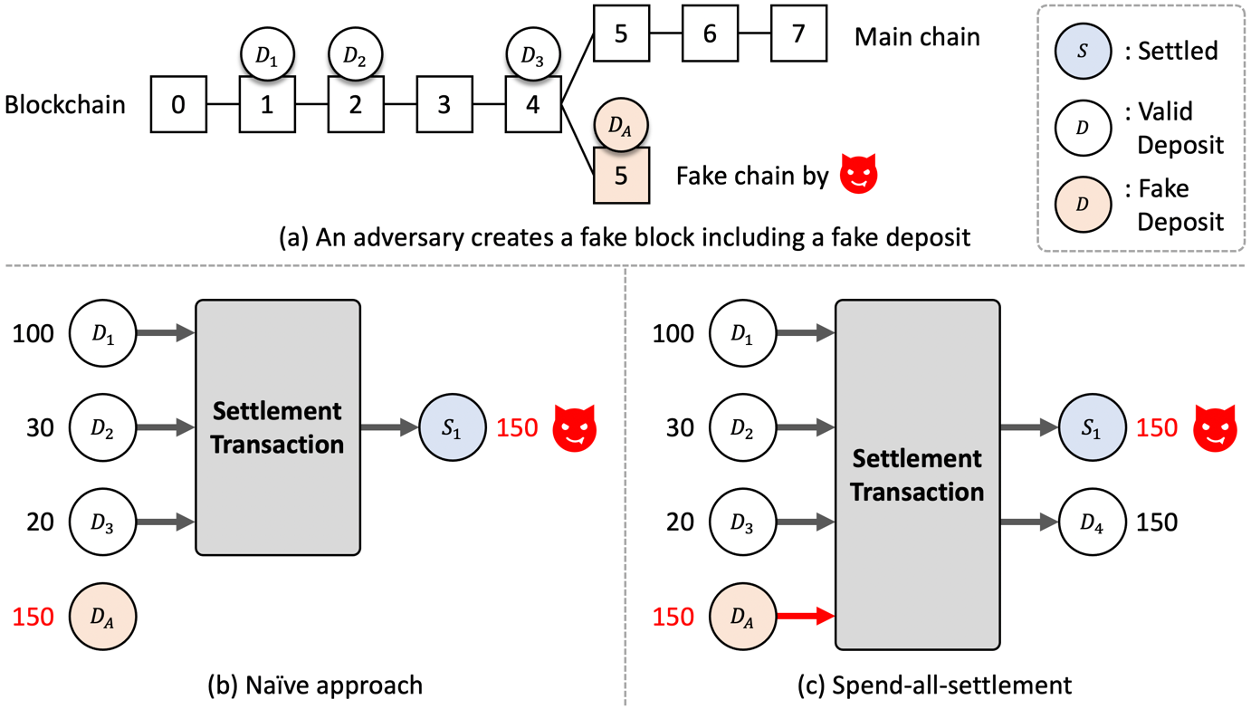}
    \caption{
        An attack scenario using fake blockchain data. 
        (a) White blocks are included in a valid main chain. 
        A red block is generated by the malicious host. 
        The host inserts the red block into RouTEE to add the fake deposit $D_A$ and tries to settle its invalid balance. 
        (b) If RouTEE picks old deposits to make a settlement transaction (naive approach), the deposit $D_A$ is not spent, which means the transaction is valid in both chains. 
        Thus, the host can broadcast the transaction to the Bitcoin network, stealing other valid deposits. 
        (c) On the contrary, if the settlement transaction contains all deposits (spend-all-settlement), it is only valid in the fake chain as there is no $D_A$ in the main chain. 
        Thus the host is not able to gain illegal benefits. 
    }
    \label{fig:spend-all-settlement}
\end{figure*}

\subsection{Fake Blockchain Data}

All balances in RouTEE are derived from on-chain deposits locked in the blockchain. 
However, as described in Fig.~\ref{fig:spend-all-settlement}, malicious hosts could generate fake blocks, which involve fake deposit transactions not actually broadcasted in the blockchain network intended to deceive RouTEE. 
Although this attack costs a lot of mining time to create blocks with adequate difficulty, it is possible because there is no time limit and adversaries can earn invalid balances in RouTEE without staking their on-chain assets.

After obtaining illegal balances, adversaries need to execute multi-hop payments or settlements to gain actual profits. 
For this reason, users, particularly payees, have to verify whether or not payers' balances are derived from valid source blocks. 
They can efficiently filter invalid balances out by setting boundary blocks since fake blocks have different hash values than users' blocks. 
However, there might be an \textit{eclipse attack} \cite{heilman2015eclipse} where adversaries who take all the network connections of a victim supply fake blockchain information to make them believe it is the correct main chain. 
To avoid eclipse attacks, users should fetch several header chains from various full nodes, compare them, and choose the reliable one according to the longest chain rule.

Still, the host can try to settle its invalid balances to steal valid on-chain deposits from other users. 
Our key idea for the spend-all-settlement is to invalidate on-chain transactions that are made based on fake blockchain data. 
As illustrated in Fig.~\ref{fig:spend-all-settlement}, we thought of a naive approach for settlements that spend not all deposits, but rather the least set of deposits starting from the oldest one to just meet a total settlement amount. 
Settlement transactions made in this way are valid transactions in the fake chain, of course, and even in the main chain because they only use valid deposits. 
Thus adversaries can broadcast them to the blockchain network and take on-chain assets, with users then unable to reclaim their assets.

If we apply the \textit{spend-all-settlement} method, on the other hand, settlement transactions include every deposit. 
So if there is at least one fake deposit in the settlement transactions, they would be rejected from the main chain, as there is no such UTXO to spend. 
To conclude, employing boundary block verification and the \textit{spend-all-settlement} prevents illegal attempts to gain invalid benefits through payments and settlements, respectively. 
Therefore, the best strategy for rational hosts is simply feeding correct blockchain data to confirm user balances and earn routing fees.

\subsection{Abortion}

Hosts can abort operating RouTEE in various ways. 
First, the host can simply shut down the machine running RouTEE. 
If SGX is turned off, all private data in SGX such as user balance states and the block header chain disappear because the secure memory space in SGX is volatile. 
However, a rational host would not forcibly shut down RouTEE, as the host does not want to abandon the pending routing fees in it. 
To terminate RouTEE normally, the host should go through a termination process that settles all users' balances to the settle addresses obtained in advance by broadcasting settlement transactions. 
The host could then take all confirmed routing fees.

Next, the host could stop feeding blockchain data into RouTEE. 
If there are deposit transactions not yet inserted into SGX, then they are bound in the blockchain forever, and there is no way to retrieve it. 
Similarly, the host could generate a fake blockchain that does not contain the user's deposit, but the host has no financial motivation to ignore deposits, losing all routing fees and maybe spending an expensive mining cost.

We might fundamentally prevent these attacks by utilizing \textit{time-locked transactions}, which make deposits retrievable after a certain amount of time. 
However, their transaction fees are more expensive than P2PKH transactions, and RouTEE must broadcast additional transactions that bring time-locked assets to RouTEE within the time limit (users may need to pay for this transaction also). 
To conclude, P2PKH transactions are appropriate for RouTEE because we are only concerned about rational adversaries and time-locked transactions are inefficient and unnecessarily expensive.

Lastly, the host might not broadcast settlement transactions. 
Pending routing fees induce the host to fulfill its duty since abandoning settlement transactions means that the host gives up confirming its pending routing fees. 
What's more, settlement transactions are sequentially included in the blockchain, as they spend the leftover deposit from the previous one. 
This means that if the host drops a settlement transaction, the host cannot broadcast any settlement transactions and cannot obtain any confirmed routing fees from that moment. 
Thus hosts are motivated to broadcast settlement transactions honestly.

\subsection{Message Abusing}

As messages between users and RouTEE are encrypted, hosts cannot find out their contents or manipulate them. 
They are not able to link a user address and the user's on-chain deposit, nor do they know the details of multi-hop payments and settlements.

Hosts may try to drop or delay messages, but it does not give them any financial benefit since operations in RouTEE bring and confirm routing fees. 
Even if they are going to censor a particular user's messages or requests for a specific operation in the face of a potential financial loss, secure sessions between them would make it impossible.

Since the minimum amount of the routing fee is determined by the hosts, they can execute a \textit{message reordering attack}, in which they extort a payer's balance by delivering the user's multi-hop payment message to RouTEE after changing the minimum routing fee amount. 
Users can easily prevent this attack by specifying their routing fee amounts in their messages. 
In case of a \textit{replay attack}, where adversaries insert the same message several times into RouTEE (e.g., execute the same multi-hop payments to get routing fees), a user's nonce field will efficiently block it.

% \clearpage

\section{Evaluation}

We implement a RouTEE prototype for Bitcoin using SGX. 
We use the Intel SGX SDK for Linux \cite{SGXSDK} and Bitcoin core libraries \cite{BitcoinCore}. 
To generate Bitcoin addresses (i.e., manager addresses), we utilize secp256k1 which is available from the Bitcoin core. 
From SGX SDK, we choose AES128-GCM to encrypt/decrypt messages between users and RouTEE and RSA digital signatures created by 3072-bit keys. 
Bitcoin employs ECDSA signatures, but, being unsuitable for RouTEE, it takes a lot of time to verify signatures.

% Table \ref{tab:user_operation_performance}
\begin{table}[t]
    \caption{RouTEE User operation performance}
    \centering
    
    \begin{tabular}{l|r|rr}\hline
         & Throughput & \multicolumn{2}{c}{Latency (ms)} \\
         Operation & (op/sec) & Idle $[99^{th} \%]$ & Busy $[99^{th} \%]$ \\\hline
         \textit{add user} & 1,222 & 42.7 [45.9] & 47.1 [50.6] \\
         \textit{add deposit} & 1,092 & 44.8 [59.3] & 48.1 [55.1] \\
         \textit{update boundary block} & 19,998 & 38.8 [41.7] & 37.4 [40.1] \\
         \textit{multi-hop payment} & 18,677 & 38.6 [48.4] & 43.2 [56.0] \\
         \textit{settlement} & 19,607 & 38.9 [43.5] & 36.5 [39.6] \\\hline
    \end{tabular}
    
    \label{tab:user_operation_performance}
\end{table}

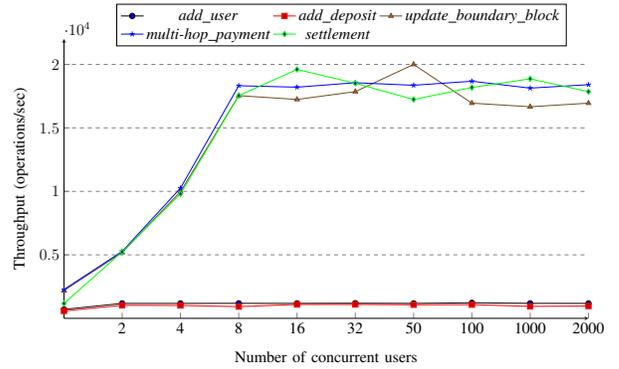
\begin{figure}[t]
\centering
    \begin{adjustbox}{width=0.9\linewidth}
    \begin{tikzpicture}
    \begin{axis}[
        axis lines=middle,
        title style={at={(0.5,0.94)},anchor=north,yshift=-0.1,font=\Large},
        height=9cm,
        width=15.5cm,
        ymin=0,
        ymax=22000,
        xmin=1,
        xmax=4500,
        xlabel = Number of concurrent users,
        ylabel = Throughput (operations/sec),
        ymajorgrids, tick align=inside,
        major grid style={dashed,draw=black!50},
        x label style={at={(axis description cs:0.5,-0.1)},anchor=north,font=\large},
        y label style={at={(axis description cs:-0.06,.5)},rotate=90,anchor=south,font=\large},
        x tick label style={font=\large},
        y tick label style={font=\large},
        legend style={at={(0.1,1.05)}, anchor=west, font=\large, legend columns=3},
        enlargelimits = false,
        xticklabels from table={data/user_ops_tps.data}{users},
        xtick=data
    ]
    
    % draw lines
    \addplot+[color=black, sharp plot] table [y=add_user,x=X]{data/user_ops_tps.data};
    \addlegendentry{\textit{add\_user}}
    
    \addplot+[sharp plot] table [y=add_deposit,x=X]{data/user_ops_tps.data};
    \addlegendentry{\textit{add\_deposit}}
    
    \addplot+[sharp plot, mark=triangle*] table [y=update_boundary_block,x=X]{data/user_ops_tps.data};
    \addlegendentry{\textit{update\_boundary\_block}}
    
    \addplot+[color=blue, sharp plot] table [y=multi-hop_payment,x=X]{data/user_ops_tps.data};
    \addlegendentry{\textit{multi-hop\_payment}}
    
    \addplot+[color=green, sharp plot] table [y=settlement,x=X]{data/user_ops_tps.data};
    \addlegendentry{\textit{settlement}}
    
    \end{axis}
    \end{tikzpicture}
    \end{adjustbox}
\caption{User operation throughput}
\label{fig:tps_graph}
\end{figure}

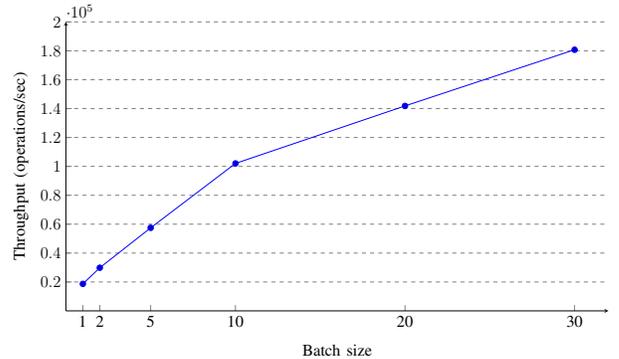
\begin{figure}[t]
\centering
    \begin{adjustbox}{width=0.9\linewidth}
    \begin{tikzpicture}
    \begin{axis}[
        axis lines=middle,
        title style={at={(0.5,0.94)},anchor=north,yshift=-0.1,font=\Large},
        height=9cm,
        width=15.5cm,
        ymin=0,
        ymax=200000,
        xmin=0,
        xmax=32,
        xlabel = Batch size,
        ylabel = Throughput (operations/sec),
        ymajorgrids, tick align=inside,
        major grid style={dashed,draw=black!50},
        x label style={at={(axis description cs:0.5,-0.1)},anchor=north,font=\large},
        y label style={at={(axis description cs:-0.06,.5)},rotate=90,anchor=south,font=\large},
        x tick label style={font=\large},
        y tick label style={font=\large},
        enlargelimits = false,
        xticklabels from table={data/payment_batch_tps.data}{users},
        xtick=data
    ]
    
    % draw lines
    \addplot+[sharp plot] table [y=tps,x=X]{data/payment_batch_tps.data};
    
    \end{axis}
    \end{tikzpicture}
    \end{adjustbox}
\caption{Batched payments throughput}
\label{fig:batched_payments_tps_graph}
\end{figure}

\subsection{Performance of User Operations}

For experiments, we implement user clients that send encrypted and signed messages to RouTEE using Python. 
We run a RouTEE program on a machine with an Intel i9-9900K CPU and 64GB of RAM running Ubuntu 16.04 LTS. 
It has 8-core/16-thread but we only use four threads.

\textbf{Throughput.} 
We define the throughput as the number of operations RouTEE can process in one second. 
In order to measure RouTEE's throughput for each user operation, we make user clients continuously send the same type of user operation messages with random parameters (e.g., user addresses, payment amounts). 
Supplying a sufficient amount of messages to find a maximum throughput value, we execute multiple client programs on a machine with an AMD Ryzen Threadripper 2990WX CPU, 128GB of RAM, and 32-core/64-thread running Ubuntu 18.04 LTS, connected with the RouTEE server over the local network. 
We also repeat this experiment with various numbers of concurrent users. 
As suggested in Fig.~\ref{fig:tps_graph}, RouTEE deals with thousands of concurrent users, maintaining its maximum transaction throughput.

As it executes dynamic allocation to register users in the user state, the \textit{add\_user} operation shows a maximum throughput of about 1,200 op/sec, which is relatively low compared to other operations. 
Similarly, the \textit{add\_deposit} operation has a maximum throughput of about 1,100 op/sec, also quite low relatively speaking, as it generates random private keys to make random manager addresses. 
However, these operations are more scalable than other payment networks when we consider that one Bitcoin block cannot include more than 3,000 transactions, meaning that they cannot open more than 3,000 channels every 10 minutes.

The \textit{update\_boundary\_block} and settlement operations show similar throughput of about 18,000 op/sec. 
Given the nature of payment networks, \textit{update\_boundary\_block} would occur more frequently than settlements. 
Even if every 100,000 users update their boundary blocks when a new block is created, it will only take 6 seconds. 
We note that there are about 15,000 public users in Lightning Network, opening about 36,000 channels as of November 2020 \cite{1ML}.

The multi-hop payment's throughput is about 18,000 op/sec. 
This is similar to existing payment systems such as VISA, which is known to be capable of dealing with about 24,000 payments \cite{VISAcard}. 
Furthermore, we can boost the throughput by batching multi-hop payments. 
Senders can make batched multi-hop payments that contain multiple receivers and payment amounts. 
We measure the throughput of batched payments and the result is shown in Fig.~\ref{fig:batched_payments_tps_graph}. 
The throughput increases as the size of the batch increases, and users can choose various batch sizes. 
For example, RouTEE achieves an approximately 30,000 op/sec throughput when the size is two, and about 180,000 op/sec when the size is 30. 
The graph is not completely linear because as the size increases, the message size also grows, costing more network overhead. 
We note that the throughput can be enhanced by employing additional hub nodes or multi threads. 
The results of the throughputs are listed in Table \ref{tab:user_operation_performance}.

\begin{table}[t]
    \caption{Light client execution time for downloading and verifying 2,016 valid block headers}
    \centering
    
    \begin{tabular}{l|rrrrr}\hline
         & \multicolumn{5}{c}{Time (sec)} \\
         & Min & Med & Avg & $99^{th} \%$ & Max \\\hline
         Network latency & 0.161 & 0.315 & 0.499 & 3.744 & 16.679 \\
         Verification time & 0.054 & 0.088 & 0.089 & 0.126 & 0.297 \\\hline
         Total & 0.216 & 0.404 & 0.588 & 3.871 & 16.977 \\\hline
    \end{tabular}
    
    \label{tab:light_client_performance}
\end{table}

\textbf{Latency.} 
We define the latency as the time between sending a message to RouTEE and receiving a response message from RouTEE. 
To measure the RouTEE's latency for each user operation in a realistic network environment, we run the client program on a t2.micro Amazon EC2 instance (RTT: 39.9 ms, Bandwidth: 31.4 Mbits). 
The client sends a message to RouTEE, waits for the response, and repeats this continuously. 
Table \ref{tab:user_operation_performance} shows the result. 
Idle means that there is no concurrent user in RouTEE, and Busy means that RouTEE is executing operations at its maximum throughput speed. 
Most operations are returned within about 50 ms, which is a very small latency, and there is almost no difference in the latency between idle and busy states.

\subsection{Performance of Host Operations}
To provide blockchain data, we run the Bitcoin core program as a full node on the same machine as RouTEE. 
Since we do not need to evaluate Bitcoin network interactions, we simply establish the Bitcoin local private network, called \textit{regnet}, and create 100 blocks which contain 2,500 randomly generated transactions each (for almost any period, the average number of transactions per block does not exceed 2,500 \cite{AvgTxNum}). 
The RouTEE program can obtain blockchain data such as headers and transactions through RPC requests to the Bitcoin core.

\textbf{Block Verification.} 
We first insert 100 block headers to measure the block header verification time (i.e., a light client's verification inside SGX). 
Then we insert 100 headers with their transactions to measure the \textit{insert\_block} operation execution time. 
The result is that it takes 18.56 ms to verify a single header and 45.16 ms for an \textit{insert\_block} operation.

\textbf{Initialization.} 
To initialize RouTEE, a host needs to feed a header chain to RouTEE. 
Based on the result above, it takes about 3.37 hours to start verifying headers from the genesis block to the latest block (about 655,000 blocks). 
If we start from the 295,000th block, which is the latest checkpoint block in the Bitcoin core \cite{BitcoinCheckpointBlock}, it takes 1.85 hours. 
Furthermore, it only takes 37.41 seconds to verify 2,016 blocks by hard-coding the block that is 2,016 blocks away from the latest block. 
It takes an additional one minute to verify 2,016 full blocks to finish initialization, calculating $\txAvgFee$. 
We note that the initialization needs to be performed only once.

\begin{figure}[t]
\centering
    \begin{adjustbox}{width=0.9\linewidth}
    \begin{tikzpicture}
    \begin{axis}[
        axis lines=middle,
        title style={at={(0.5,0.94)},anchor=north,yshift=-0.1,font=\Large},
        height=9cm,
        width=15.5cm,
        ymin=0,
        ymax=10,
        xmin=1,
        xmax=2500,
        xlabel = Number of transaction inputs,
        ylabel = Time (s),
        ymajorgrids, tick align=inside,
        major grid style={dashed,draw=black!50},
        x label style={at={(axis description cs:0.5,-0.1)},anchor=north,font=\large},
        y label style={at={(axis description cs:-0.06,.5)},rotate=90,anchor=south,font=\large},
        x tick label style={font=\large},
        y tick label style={font=\large},
        legend style={at={(0.1,1.0)}, anchor=west, font=\large, legend columns=4},
        enlargelimits = false,
        xticklabels from table={data/settle_tx_gen_time.data}{intputNum},
        xtick=data
    ]
    
    % draw lines
    \addplot+[sharp plot] table [y=outputNum500,x=X]{data/settle_tx_gen_time.data};
    \addlegendentry{500 outputs}
    
    \addplot+[sharp plot] table [y=outputNum1000,x=X]{data/settle_tx_gen_time.data};
    \addlegendentry{1000 outputs}
    
    \addplot+[sharp plot, mark=triangle*] table [y=outputNum1500,x=X]{data/settle_tx_gen_time.data};
    \addlegendentry{1500 outputs}
    
    \addplot+[color=green, sharp plot] table [y=outputNum2000,x=X]{data/settle_tx_gen_time.data};
    \addlegendentry{2000 outputs}
    
    \end{axis}
    \end{tikzpicture}
    \end{adjustbox}
\caption{Settlement transaction generation time}
\label{fig:settle_tx_gen_time}
\end{figure}
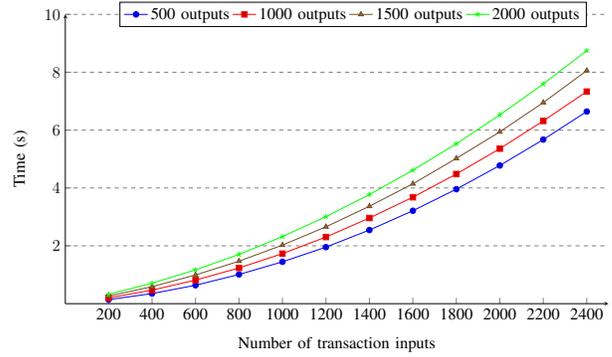

\textbf{Make Settlement Transaction.} 
To measure the settlement transaction generation time, we repeatedly create settlement transactions with various numbers of inputs and outputs within RouTEE's SGX. 
Fig.~\ref{fig:settle_tx_gen_time} shows the result. 
The execution time increases as the number of transaction inputs increases, and it takes a little more time to deal with more outputs. 
This is because most of the time is spent making signatures for inputs and hashing data several times, and the hashing process time grows quadratically as the inputs increase.

The number of inputs equals the number of deposits, and the number of outputs equals the number of settle request users (plus one for a leftover deposit). 
If RouTEE owns 2,000 deposits and 2,000 users request settlement, it only takes 6.52 seconds to generate the settlement transaction, with the transaction's size being roughly 364 KB. 
As Bitcoin's block size limit is about 1 MB, the transaction is sufficiently small to be included in the block, meaning that RouTEE can settle thousands of users simultaneously within a few seconds.

% Fig.~\ref{fig:settle_tx_gen_time}

\subsection{Performance of Light Client}

We also measure how long it takes to download header chains through a light client. 
We choose Electrum \cite{Electrum} which is one of the most popular light clients for Bitcoin. 
Electrum batches 2,016 block headers and sends header requests to various full nodes. 
We run Electrum on a t2.micro Amazon EC2 instance and note that it has very limited network resources. 
Electrum downloads a header chain from the genesis block to the latest block, and we repeat this process 100 times.

Table \ref{tab:light_client_performance} shows the result. 
When Electrum requests to the close full node, it only takes 0.161 seconds to download 2,016 headers.
Verifying these headers can be finished within 0.054 seconds, meaning that users need only 1.16 minutes to get the whole verified header chain in the best case. 
Most of the execution time is spent waiting for the download to be finished, and it varies depending on which full node is connected. 
When Electrum connects to the node far away from it, it takes 16.977 seconds in total, indicating that it requires at most 1.53 hours to download all headers. 
However, we notice that after downloading it once, users can easily catch up to the latest block within about 17 seconds every two weeks (i.e., 2,016 blocks). 
We also stress that users only need to download headers, meaning that they only require about 55 MB to store all headers.

\subsection{Comparison of Multi-hop Payment Performances}

We compare RouTEE to Lightning Network, which is the state-of-the-art payment network and has an open-sourced implementation called Lightning Network Daemon (LND) \cite{LND}. 
However, it is difficult to compare them objectively since they have quite different network topologies and protocols. 
In Lightning Network, the multi-hop payment results would vary greatly depending on the network state, routing users, and payment amounts. 
Unfortunately, there is no public data on network topologies or payment history in payment networks, so it is hard to measure the realistic performance of Lightning Network. 
By contrast, RouTEE's multi-hop payments never fail if participants are rational. 
Thus, if we can measure their performance in realistic environments, RouTEE will outperform Lightning Network.

For this reason, we measure the performance of the 1-hop payment in Lightning Network because it is a lot similar to the multi-hop payment in RouTEE without updating the receiver's boundary block (i.e., the sender simply sends the multi-hop payment message). 
In Lightning Network, receivers initiate payments, forwarding their invoices to senders. 
Then senders decode invoices and send response messages to receivers, which is the end of 1-hop payments. 
To make the equivalent environment, the sender client runs on the machine that executed the RouTEE program, and the receiver client runs on the machine that executed the client program, measuring the maximum invoice decoding throughput in a similar manner. 
As a result, it takes 28.033 seconds to process 10,000 invoices, which means that its throughput is about 360 payments/sec with a latency of 165 ms. 
Considering RouTEE's payment throughput (i.e., 18,677 payments/sec with a latency of about 40 ms from Table \ref{tab:user_operation_performance}), Lightning Network nodes are not suitable for acting as a hub node.

% \newpage

\section{Discussion} 
In this section, we consider security issues in more detail and describe possible extensions to protect RouTEE even from various kinds of irrational adversaries to enhance its security.

\subsection{Obtaining Valid Blockchain Data}

It is hard to determine whether this blockchain data is absolutely valid or not. 
We simply find a chain that looks relatively more valid than the other ones (e.g., by length of chain and block difficulty). 
Judging the chain's validity in the SGX is even more difficult as the host may restrict input data into SGX. 
Also, users might receive invalid chain data, if they are victims of eclipse attacks or carelessly download only from a single malicious node. 
In short, there might be the case that a malicious host profits from fake data.

To protect these kinds of naive users, we can the leverage statistical characteristics of block mining \cite{bentov2019tesseract}. 
In Bitcoin, a block is mined every 10 minutes on average, and the block interval time between two blocks is a random variable that has an exponential distribution. 
Adversaries with less mining power than the main chain's miners will have difficulty keeping up with this mining speed. 
Thus it is possible to detect weird blocks inserted too slowly, by measuring block interval times with a trusted relative timer inside SGX. 
However, this may yield false-positive results due to innocent but unlucky blocks, which result in deferring block acceptance for a certain amount of time (e.g., for 120 blocks).

There is another approach \cite{zhang2016town} which brings authenticated data into SGX from trustworthy web sites. 
Nowadays, many web sites \cite{Blockchaindotcom, BTCdotcom} provide blockchain-related data in real-time. 
RouTEE can therefore obtain block headers and transactions without hosts. 
In order to reduce the risk of website hacking, RouTEE should interact with various websites and decide which data is correct by a majority vote. 
This could be combined with the above method to mitigate false-positive results.

Breaking this system is quite impractical since the adversary should have mining power equivalent to all the Bitcoin miners and be able to manipulate several websites at once. 
Thus, users who trust data obtained in this manner no longer need to run light clients and can set their boundary block as the latest block inside SGX. 
This makes RouTEE more convenient for users.

\subsection{Crash Fault Tolerance}

There might be unintentional software or hardware faults so we need to save internal data in SGX periodically to restart RouTEE with the same state. 
SGX supports a sealing/unsealing feature \cite{anati2013innovative} that allows data to be securely stored and loaded within SGX only. 
Before utilizing this feature, we must prevent \textit{roll-back attacks}, where an adversary gives previous data and loads it as if this were the latest one, thereby breaking the integrity of SGX. 
To do so, SGX enables implementing hardware monotonic counters with the non-volatile memory in SGX \cite{SGXSDK} in order to track and trace the current state.

However, in the case of RouTEE, the state could be changed very frequently. 
Thus, the monotonic counter in SGX might not be suitable for RouTEE, because it has a limited performance and would wear out quickly after about 1 million writes \cite{matetic2017rote}. 
To overcome this limitation, RouTEE could batch operations to reduce the number of writes to the counter. 
In addition, there is other research \cite{matetic2017rote, strackx2016ariadne} that could enhance the data integrity for SGX.

\subsection{Network Failure}

Considering unexpected long-term network failures between users and RouTEE, we could design RouTEE to be able to settle users' channels, when their balances have not changed for a certain period of time (e.g., for 1,000 blocks) in order to guarantee their assets. 
In a different approach, we could leverage the underlying blockchain to deliver messages into RouTEE against the network failure and even indiscriminate censorship, similar to another framework using a TEE \cite{das2019fastkitten}. 
In Bitcoin, users are able to write arbitrary data to the blockchain through the OP\_RETURN instructions of the script language. 
This means that RouTEE could receive encrypted messages through on-chain transactions.
It is possible to block even these messages by generating a fake chain that does not contain them, but it causes far more losses (i.e., block mining costs, abandoning routing fees) than benefits.

% \newpage

\section{Related Work} 

\textbf{Payment networks.} 
There has been a lot of research done on payment networks.
Duplex Micropayment Channels \cite{decker2015fast} utilize time-locks to ensure that the latest transaction can be broadcast first. 
Lightning Network \cite{poon2016bitcoin} and Raiden \cite{RaidenNetwork} are payment networks based on Bitcoin and Ethereum respectively.
Flare \cite{prihodko2016flare} proposes an optimized routing path searching algorithm for Lightning Network. 
Revive \cite{khalil2017revive} rebalances payment channels to re-fund depleted channels without broadcasting on-chain transactions. 
Perun \cite{dziembowski2019perun} offers a new method called \textit{virtual payment channel}, which makes the routing more efficient but requires Turing-complete smart contracts. 
Intermediate users can be virtual payment hubs, but they would need a lot of channels that have sufficient balances to create virtual channels, and their balances would be locked until virtual channels are closed. 
Sprites \cite{miller2019sprites} reduces collateral locking time during multi-hop payments and can partially add and withdraw balances.
Pisa \cite{mccorry2019pisa} allows users to go off-line for a longer period of time by employing a third party called a \textit{custodian} who monitors blockchains on their behalf. 
Fulgor and Rayo \cite{malavolta2017concurrency} are new protocols that reinforce privacy and concurrency for payment networks. 
Anonymous multi-hop locks (AMHLs) \cite{malavolta2019anonymous} enable privacy-preserving multi-hop payments.

\textbf{Trusted execution environments with blockchains.} 
Recent studies have solved various problems in blockchains through TEEs.
TownCrier \cite{zhang2016town} feeds authenticated data to smart contracts, overcoming the oracle problem.
Obscuro \cite{tran2018obscuro} is a secure mixer platform for Bitcoin to enhance the anonymity of transactions.
FastKitten \cite{das2019fastkitten} allows smart contracts to be executed over blockchains without complex languages. Also, users can efficiently interact with smart contracts by off-chain transactions. 
Tesseract \cite{bentov2019tesseract} is a secure cryptocurrency exchange that enables assets from different blockchains to be swapped. 
Ekiden \cite{cheng2019ekiden} makes smart contracts more confidential and efficient. 
Bite \cite{matetic2019bite} protects the privacy of light clients from other full nodes. 
Teechan \cite{lind2016teechan} and Teechain \cite{lind2019teechain} establish secure payment channels. 
To the best of our knowledge, Teechain is the most relevant to our work. 
However, it requires every user to have a TEE. 
Users also have to check that their counterparty's deposit transaction is included in the blockchain when they want to allocate their deposits to their channels. 
In addition, Teechain supports multi-hop payments but their mechanism is fundamentally the same as the existing payment network's method, implying that it has the same limitations such as a high probability of failure and struggling to track the network state to find payment paths.

\section{Conclusion} 

Payment networks allow more payment transactions to be handled on behalf of blockchains with low scalability, but they are still inefficient and have many limitations. 
Users have a hard time managing their channels and multi-hop payments are difficult to achieve. 
We present RouTEE, a new payment network with a secure routing hub that fully leverages a star network topology and a TEE. 
Offering payment channels with high liquidity, it supports multi-hop payments that do not fail in normal situations. 
We consider adversaries that have control of the RouTEE platform and analyze the security of our system, proving that following the protocol honestly is the best strategy for rational hosts. 
We also evaluate RouTEE and demonstrate that RouTEE provides a practical and scalable payment network.

\bibliographystyle{IEEEtran}
\bibliography{reference}

\begin{thebibliography}{10}
\providecommand{\url}[1]{#1}
\csname url@samestyle\endcsname
\providecommand{\newblock}{\relax}
\providecommand{\bibinfo}[2]{#2}
\providecommand{\BIBentrySTDinterwordspacing}{\spaceskip=0pt\relax}
\providecommand{\BIBentryALTinterwordstretchfactor}{4}
\providecommand{\BIBentryALTinterwordspacing}{\spaceskip=\fontdimen2\font plus
\BIBentryALTinterwordstretchfactor\fontdimen3\font minus \fontdimen4\font\relax}
\providecommand{\BIBforeignlanguage}[2]{{%
\expandafter\ifx\csname l@#1\endcsname\relax
\typeout{** WARNING: IEEEtran.bst: No hyphenation pattern has been}%
\typeout{** loaded for the language `#1'. Using the pattern for}%
\typeout{** the default language instead.}%
\else
\language=\csname l@#1\endcsname
\fi
#2}}
\providecommand{\BIBdecl}{\relax}
\BIBdecl

\bibitem{nakamoto2008bitcoin}
\BIBentryALTinterwordspacing
S.~Nakamoto. Bitcoin: A peer-to-peer electronic cash system. [Online]. Available: \url{https://www.bitcoin.org/bitcoin.pdf}
\BIBentrySTDinterwordspacing

\bibitem{wood2014ethereum}
G.~Wood \emph{et~al.}, ``Ethereum: A secure decentralised generalised transaction ledger,'' \emph{Ethereum project yellow paper}, vol. 151, no. 2014, pp. 1--32, 2014.

\bibitem{gervais2016security}
A.~Gervais, G.~O. Karame, K.~W{\"u}st, V.~Glykantzis, H.~Ritzdorf, and S.~Capkun, ``On the security and performance of proof of work blockchains,'' in \emph{Proceedings of the 2016 ACM SIGSAC conference on computer and communications security}, 2016, pp. 3--16.

\bibitem{poon2016bitcoin}
J.~Poon and T.~Dryja, ``The bitcoin lightning network: Scalable off-chain instant payments,'' 2016.

\bibitem{RaidenNetwork}
\BIBentryALTinterwordspacing
Raiden network. [Online]. Available: \url{https://raiden.network/}
\BIBentrySTDinterwordspacing

\bibitem{avarikioti2020ride}
Z.~Avarikioti, L.~Heimbach, Y.~Wang, and R.~Wattenhofer, ``Ride the lightning: The game theory of payment channels,'' in \emph{International Conference on Financial Cryptography and Data Security}.\hskip 1em plus 0.5em minus 0.4em\relax Springer, 2020, pp. 264--283.

\bibitem{mckeen2013innovative}
F.~McKeen, I.~Alexandrovich, A.~Berenzon, C.~V. Rozas, H.~Shafi, V.~Shanbhogue, and U.~R. Savagaonkar, ``Innovative instructions and software model for isolated execution.'' \emph{Hasp@ isca}, vol.~10, no.~1, 2013.

\bibitem{costan2016intel}
V.~Costan and S.~Devadas, ``Intel sgx explained.'' \emph{IACR Cryptol. ePrint Arch.}, vol. 2016, no.~86, pp. 1--118, 2016.

\bibitem{dwork1992pricing}
C.~Dwork and M.~Naor, ``Pricing via processing or combatting junk mail,'' in \emph{Annual International Cryptology Conference}.\hskip 1em plus 0.5em minus 0.4em\relax Springer, 1992, pp. 139--147.

\bibitem{seres2020topological}
I.~A. Seres, L.~Guly{\'a}s, D.~A. Nagy, and P.~Burcsi, ``Topological analysis of bitcoin’s lightning network,'' in \emph{Mathematical Research for Blockchain Economy}.\hskip 1em plus 0.5em minus 0.4em\relax Springer, 2020, pp. 1--12.

\bibitem{herrera2019difficulty}
J.~Herrera-Joancomart{\'\i}, G.~Navarro-Arribas, A.~Ranchal-Pedrosa, C.~P{\'e}rez-Sol{\`a}, and J.~Garcia-Alfaro, ``On the difficulty of hiding the balance of lightning network channels,'' in \emph{Proceedings of the 2019 ACM Asia Conference on Computer and Communications Security}, 2019, pp. 602--612.

\bibitem{kappos2020empirical}
G.~Kappos, H.~Yousaf, A.~Piotrowska, S.~Kanjalkar, S.~Delgado-Segura, A.~Miller, and S.~Meiklejohn, ``An empirical analysis of privacy in the lightning network,'' \emph{arXiv preprint arXiv:2003.12470}, 2020.

\bibitem{malavolta2019anonymous}
G.~Malavolta, P.~Moreno-Sanchez, C.~Schneidewind, A.~Kate, and M.~Maffei, ``Anonymous multi-hop locks for blockchain scalability and interoperability.'' in \emph{NDSS}, 2019.

\bibitem{perez2020lockdown}
C.~P{\'e}rez-Sol{\`a}, A.~Ranchal-Pedrosa, J.~Herrera-Joancomart{\'\i}, G.~Navarro-Arribas, and J.~Garcia-Alfaro, ``Lockdown: Balance availability attack against lightning network channels,'' in \emph{International Conference on Financial Cryptography and Data Security}.\hskip 1em plus 0.5em minus 0.4em\relax Springer, 2020, pp. 245--263.

\bibitem{rohrer2019discharged}
E.~Rohrer, J.~Malliaris, and F.~Tschorsch, ``Discharged payment channels: Quantifying the lightning network's resilience to topology-based attacks,'' in \emph{2019 IEEE European Symposium on Security and Privacy Workshops (EuroS\&PW)}.\hskip 1em plus 0.5em minus 0.4em\relax IEEE, 2019, pp. 347--356.

\bibitem{anati2013innovative}
I.~Anati, S.~Gueron, S.~Johnson, and V.~Scarlata, ``Innovative technology for cpu based attestation and sealing,'' in \emph{Proceedings of the 2nd international workshop on hardware and architectural support for security and privacy}, vol.~13.\hskip 1em plus 0.5em minus 0.4em\relax Citeseer, 2013, p.~7.

\bibitem{brasser2017software}
F.~Brasser, U.~M{\"u}ller, A.~Dmitrienko, K.~Kostiainen, S.~Capkun, and A.-R. Sadeghi, ``Software grand exposure:$\{$SGX$\}$ cache attacks are practical,'' in \emph{11th $\{$USENIX$\}$ Workshop on Offensive Technologies ($\{$WOOT$\}$ 17)}, 2017.

\bibitem{lee2017hacking}
J.~Lee, J.~Jang, Y.~Jang, N.~Kwak, Y.~Choi, C.~Choi, T.~Kim, M.~Peinado, and B.~B. Kang, ``Hacking in darkness: Return-oriented programming against secure enclaves,'' in \emph{26th $\{$USENIX$\}$ Security Symposium ($\{$USENIX$\}$ Security 17)}, 2017, pp. 523--539.

\bibitem{van2018foreshadow}
J.~Van~Bulck, M.~Minkin, O.~Weisse, D.~Genkin, B.~Kasikci, F.~Piessens, M.~Silberstein, T.~F. Wenisch, Y.~Yarom, and R.~Strackx, ``Foreshadow: Extracting the keys to the intel $\{$SGX$\}$ kingdom with transient out-of-order execution,'' in \emph{27th $\{$USENIX$\}$ Security Symposium ($\{$USENIX$\}$ Security 18)}, 2018, pp. 991--1008.

\bibitem{biondo2018guard}
A.~Biondo, M.~Conti, L.~Davi, T.~Frassetto, and A.-R. Sadeghi, ``The guard's dilemma: Efficient code-reuse attacks against intel $\{$SGX$\}$,'' in \emph{27th $\{$USENIX$\}$ Security Symposium ($\{$USENIX$\}$ Security 18)}, 2018, pp. 1213--1227.

\bibitem{chen2019sgxpectre}
G.~Chen, S.~Chen, Y.~Xiao, Y.~Zhang, Z.~Lin, and T.~H. Lai, ``Sgxpectre: Stealing intel secrets from sgx enclaves via speculative execution,'' in \emph{2019 IEEE European Symposium on Security and Privacy (EuroS\&P)}.\hskip 1em plus 0.5em minus 0.4em\relax IEEE, 2019, pp. 142--157.

\bibitem{seo2017sgx}
J.~Seo, B.~Lee, S.~M. Kim, M.-W. Shih, I.~Shin, D.~Han, and T.~Kim, ``Sgx-shield: Enabling address space layout randomization for sgx programs.'' in \emph{NDSS}, 2017.

\bibitem{shih2017t}
M.-W. Shih, S.~Lee, T.~Kim, and M.~Peinado, ``T-sgx: Eradicating controlled-channel attacks against enclave programs.'' in \emph{NDSS}, 2017.

\bibitem{chen2017detecting}
S.~Chen, X.~Zhang, M.~K. Reiter, and Y.~Zhang, ``Detecting privileged side-channel attacks in shielded execution with d{\'e}j{\'a} vu,'' in \emph{Proceedings of the 2017 ACM on Asia Conference on Computer and Communications Security}, 2017, pp. 7--18.

\bibitem{oleksenko2018varys}
O.~Oleksenko, B.~Trach, R.~Krahn, M.~Silberstein, and C.~Fetzer, ``Varys: Protecting $\{$SGX$\}$ enclaves from practical side-channel attacks,'' in \emph{2018 $\{$Usenix$\}$ Annual Technical Conference ($\{$USENIX$\}$$\{$ATC$\}$ 18)}, 2018, pp. 227--240.

\bibitem{ahmad2019obfuscuro}
A.~Ahmad, B.~Joe, Y.~Xiao, Y.~Zhang, I.~Shin, and B.~Lee, ``Obfuscuro: A commodity obfuscation engine on intel sgx.'' in \emph{NDSS}, 2019.

\bibitem{alves2004trustzone}
T.~Alves, ``Trustzone: Integrated hardware and software security,'' \emph{White paper}, 2004.

\bibitem{dong2017betrayal}
C.~Dong, Y.~Wang, A.~Aldweesh, P.~McCorry, and A.~van Moorsel, ``Betrayal, distrust, and rationality: Smart counter-collusion contracts for verifiable cloud computing,'' in \emph{Proceedings of the 2017 ACM SIGSAC Conference on Computer and Communications Security}, 2017, pp. 211--227.

\bibitem{shostak1982byzantine}
R.~Shostak, M.~Pease, and L.~Lamport, ``The byzantine generals problem,'' \emph{ACM Transactions on Programming Languages and Systems}, vol.~4, no.~3, pp. 382--401, 1982.

\bibitem{heilman2015eclipse}
E.~Heilman, A.~Kendler, A.~Zohar, and S.~Goldberg, ``Eclipse attacks on bitcoin’s peer-to-peer network,'' in \emph{24th $\{$USENIX$\}$ Security Symposium ($\{$USENIX$\}$ Security 15)}, 2015, pp. 129--144.

\bibitem{SGXSDK}
\BIBentryALTinterwordspacing
{Intel Software Guard Extensions SDK Developer Reference for Linux OS}. [Online]. Available: \url{https://download.01.org/intel-sgx/linux-2.1.3/docs/Intel_SGX_Developer_Reference_Linux_2.1.3_Open_Source.pdf}
\BIBentrySTDinterwordspacing

\bibitem{BitcoinCore}
\BIBentryALTinterwordspacing
{The Bitcoin Community}. {Bitcoin Core version 0.13.1 released}. [Online]. Available: \url{https://bitcoin.org/bin/bitcoin-core-0.13.1/}
\BIBentrySTDinterwordspacing

\bibitem{1ML}
\BIBentryALTinterwordspacing
{1ML}. {1ML: Lightning Network Search and Analysis Engine}. [Online]. Available: \url{https://1ml.com/}
\BIBentrySTDinterwordspacing

\bibitem{VISAcard}
\BIBentryALTinterwordspacing
{VISA}. {Visa's transactions per second}. [Online]. Available: \url{https://usa.visa.com/run-your-business/small-business-tools/retail.html}
\BIBentrySTDinterwordspacing

\bibitem{AvgTxNum}
\BIBentryALTinterwordspacing
{Blockchain.com}. {Average transactions per block}. [Online]. Available: \url{https://www.blockchain.com/charts/n-transactions-per-block}
\BIBentrySTDinterwordspacing

\bibitem{BitcoinCheckpointBlock}
\BIBentryALTinterwordspacing
{Bitcoin Core}. {The latest checkpoint block}. [Online]. Available: \url{https://github.com/bitcoin/bitcoin/blob/master/src/chainparams.cpp#L160}
\BIBentrySTDinterwordspacing

\bibitem{Electrum}
\BIBentryALTinterwordspacing
{Electrum}. {Electrum: Bitcoin Wallet}. [Online]. Available: \url{https://electrum.org/#home}
\BIBentrySTDinterwordspacing

\bibitem{LND}
\BIBentryALTinterwordspacing
{Lightning Network Community}. {Lightning Network Daemon}. [Online]. Available: \url{https://github.com/lightningnetwork/lnd}
\BIBentrySTDinterwordspacing

\bibitem{bentov2019tesseract}
I.~Bentov, Y.~Ji, F.~Zhang, L.~Breidenbach, P.~Daian, and A.~Juels, ``Tesseract: Real-time cryptocurrency exchange using trusted hardware,'' in \emph{Proceedings of the 2019 ACM SIGSAC Conference on Computer and Communications Security}, 2019, pp. 1521--1538.

\bibitem{zhang2016town}
F.~Zhang, E.~Cecchetti, K.~Croman, A.~Juels, and E.~Shi, ``Town crier: An authenticated data feed for smart contracts,'' in \emph{Proceedings of the 2016 aCM sIGSAC conference on computer and communications security}, 2016, pp. 270--282.

\bibitem{Blockchaindotcom}
\BIBentryALTinterwordspacing
{Blockchain.com: Blockchain Explorer}. [Online]. Available: \url{https://www.blockchain.com/explorer}
\BIBentrySTDinterwordspacing

\bibitem{BTCdotcom}
\BIBentryALTinterwordspacing
{BTC.com: Bitcoin Explorer}. [Online]. Available: \url{https://btc.com/}
\BIBentrySTDinterwordspacing

\bibitem{matetic2017rote}
S.~Matetic, M.~Ahmed, K.~Kostiainen, A.~Dhar, D.~Sommer, A.~Gervais, A.~Juels, and S.~Capkun, ``$\{$ROTE$\}$: Rollback protection for trusted execution,'' in \emph{26th $\{$USENIX$\}$ Security Symposium ($\{$USENIX$\}$ Security 17)}, 2017, pp. 1289--1306.

\bibitem{strackx2016ariadne}
R.~Strackx and F.~Piessens, ``Ariadne: A minimal approach to state continuity,'' in \emph{25th $\{$USENIX$\}$ Security Symposium ($\{$USENIX$\}$ Security 16)}, 2016, pp. 875--892.

\bibitem{das2019fastkitten}
P.~Das, L.~Eckey, T.~Frassetto, D.~Gens, K.~Host{\'a}kov{\'a}, P.~Jauernig, S.~Faust, and A.-R. Sadeghi, ``Fastkitten: Practical smart contracts on bitcoin,'' in \emph{28th $\{$USENIX$\}$ Security Symposium ($\{$USENIX$\}$ Security 19)}, 2019, pp. 801--818.

\bibitem{decker2015fast}
C.~Decker and R.~Wattenhofer, ``A fast and scalable payment network with bitcoin duplex micropayment channels,'' in \emph{Symposium on Self-Stabilizing Systems}.\hskip 1em plus 0.5em minus 0.4em\relax Springer, 2015, pp. 3--18.

\bibitem{prihodko2016flare}
P.~Prihodko, S.~Zhigulin, M.~Sahno, A.~Ostrovskiy, and O.~Osuntokun, ``Flare: An approach to routing in lightning network,'' \emph{White Paper}, 2016.

\bibitem{khalil2017revive}
R.~Khalil and A.~Gervais, ``Revive: Rebalancing off-blockchain payment networks,'' in \emph{Proceedings of the 2017 ACM SIGSAC Conference on Computer and Communications Security}, 2017, pp. 439--453.

\bibitem{dziembowski2019perun}
S.~Dziembowski, L.~Eckey, S.~Faust, and D.~Malinowski, ``Perun: Virtual payment hubs over cryptocurrencies,'' in \emph{2019 IEEE Symposium on Security and Privacy (SP)}.\hskip 1em plus 0.5em minus 0.4em\relax IEEE, 2019, pp. 106--123.

\bibitem{miller2019sprites}
A.~Miller, I.~Bentov, S.~Bakshi, R.~Kumaresan, and P.~McCorry, ``Sprites and state channels: Payment networks that go faster than lightning,'' in \emph{International Conference on Financial Cryptography and Data Security}.\hskip 1em plus 0.5em minus 0.4em\relax Springer, 2019, pp. 508--526.

\bibitem{mccorry2019pisa}
P.~McCorry, S.~Bakshi, I.~Bentov, S.~Meiklejohn, and A.~Miller, ``Pisa: Arbitration outsourcing for state channels,'' in \emph{Proceedings of the 1st ACM Conference on Advances in Financial Technologies}, 2019, pp. 16--30.

\bibitem{malavolta2017concurrency}
G.~Malavolta, P.~Moreno-Sanchez, A.~Kate, M.~Maffei, and S.~Ravi, ``Concurrency and privacy with payment-channel networks,'' in \emph{Proceedings of the 2017 ACM SIGSAC Conference on Computer and Communications Security}, 2017, pp. 455--471.

\bibitem{tran2018obscuro}
M.~Tran, L.~Luu, M.~S. Kang, I.~Bentov, and P.~Saxena, ``Obscuro: A bitcoin mixer using trusted execution environments,'' in \emph{Proceedings of the 34th Annual Computer Security Applications Conference}, 2018, pp. 692--701.

\bibitem{cheng2019ekiden}
R.~Cheng, F.~Zhang, J.~Kos, W.~He, N.~Hynes, N.~Johnson, A.~Juels, A.~Miller, and D.~Song, ``Ekiden: A platform for confidentiality-preserving, trustworthy, and performant smart contracts,'' in \emph{2019 IEEE European Symposium on Security and Privacy (EuroS\&P)}.\hskip 1em plus 0.5em minus 0.4em\relax IEEE, 2019, pp. 185--200.

\bibitem{matetic2019bite}
S.~Matetic, K.~W{\"u}st, M.~Schneider, K.~Kostiainen, G.~Karame, and S.~Capkun, ``$\{$BITE$\}$: Bitcoin lightweight client privacy using trusted execution,'' in \emph{28th $\{$USENIX$\}$ Security Symposium ($\{$USENIX$\}$ Security 19)}, 2019, pp. 783--800.

\bibitem{lind2016teechan}
J.~Lind, I.~Eyal, P.~Pietzuch, and E.~G. Sirer, ``Teechan: Payment channels using trusted execution environments,'' \emph{arXiv preprint arXiv:1612.07766}, 2016.

\bibitem{lind2019teechain}
J.~Lind, O.~Naor, I.~Eyal, F.~Kelbert, E.~G. Sirer, and P.~Pietzuch, ``Teechain: a secure payment network with asynchronous blockchain access,'' in \emph{Proceedings of the 27th ACM Symposium on Operating Systems Principles}, 2019, pp. 63--79.

\end{thebibliography}

\end{document}